\documentclass[12pt,preprint]{aastex}

\slugcomment{To appear in ApJ}

\shorttitle{Mid-IR Observations of Earth}
\shortauthors{Hearty et al.}

\begin{document}


\title{Mid-Infrared Properties of Disk Averaged Observations of Earth with AIRS}


\author{Thomas Hearty}
\affil{Goddard Space Flight Center/Wyle Information Systems, Mailstop 610.2, Greenbelt, MD 20771}
\email{Thomas.J.Hearty@nasa.gov}

\and

\author{Inseok Song}
\affil{Department of Physics and Astronomy, University of Georgia, Athens, GA 30602-2451}

\and

\author{Sam Kim}
\affil{Department of Physics and Astronomy, University of California,
Irvine, CA 92697}

\and

\author{Giovanna Tinetti}
\affil{Department of Physics and Astronomy, STFC/University College London,
Gower Street, London WC1E 6BT, UK}




\begin{abstract}
We have investigated mid-infrared spectra of Earth obtained by the
Atmospheric Infrared Sounder (AIRS) instrument on-board the AQUA spacecraft to
explore the characteristics that may someday be observed in extrasolar terrestrial planets.
We have used the AIRS infrared (R $\sim$ 1200; 3.75-15.4 microns) spectra
to construct directly-observed high-resolution
spectra of the only known life bearing planet, Earth.  
The AIRS spectra are the first
such spectra that span the seasons.
We investigate the
rotational and seasonal spectral variations
that would arise due to varying cloud amount and viewing geometry
and we explore what signatures may be observable in the mid-infrared
by the next generation of
telescopes capable of observing extrasolar terrestrial planets.

\end{abstract}





\section{Introduction}

Transit observations
have recently been obtained for giant planets orbiting stars other than the sun
\citep{deminga,charbonneaua,charbonneaub,richardson07,swain08,tin07} and there are plans
for telescopes [e.g., TPFs (Terrestrial Planet Finders), JWST
(James Webb Space Telescope), E-ELT (European Extremely large telescope)]
that will be capable of observing
terrestrial planets around other stars.
Although the terrestrial planets around other stars likely have a
broader range of characteristics than Earth or the other
terrestrial planets in our solar system, so far, Earth is the only
planet we know of that harbors life.
Therefore it is necessary to
explore the full range of spectral signatures and variability which can
be observed for Earth, the only known life bearing planet, so that we can
prepare for and understand the future observations of extrasolar
terrestrial planets.

The Atmospheric Infrared Sounder\footnote{see, http://disc.gsfc.nasa.gov/AIRS/documentation.shtml} (AIRS) on-board the AQUA spacecraft
is an excellent tool with which to observe Earth.
The AIRS obtains 2,916,000 spectra of
Earth every day with 2378 Infrared channels in the wavelength
range from 3.75 ${\rm \mu m}$ to 15.4 ${\rm \mu m}$
($\lambda/\delta \lambda \sim 1200$) and 4 visible near-infrared channels
(0.41--0.44 ${\rm \mu m}$; 0.58--0.68 ${\rm \mu m}$;
0.71--0.92 ${\rm mu m}$; 0.49--0.94 ${\rm \mu m}$).
The spacecraft has
a polar Sun synchronous orbit with a 1:30 PM ascending equator
crossing time.  The spectra are obtained as the AIRS scans
across its path from $\sim$ $-$49$^{\circ}$ to $+$49$^{\circ}$ about nadir
from an altitude of $\sim$ 700 km.
Thus, it is able to observe the entire Earth from
space and provide a rich set of spectra consisting of day, night,
land, ocean views at all latitudes.  The radiometric calibration of AIRS
has an uncertainty estimate of less than 0.2\% in the mid-infrared
and has been stable since launch \citep{ols07,aum06}.

\begin{figure}
\begin{center}
\includegraphics[width=.4\textwidth]{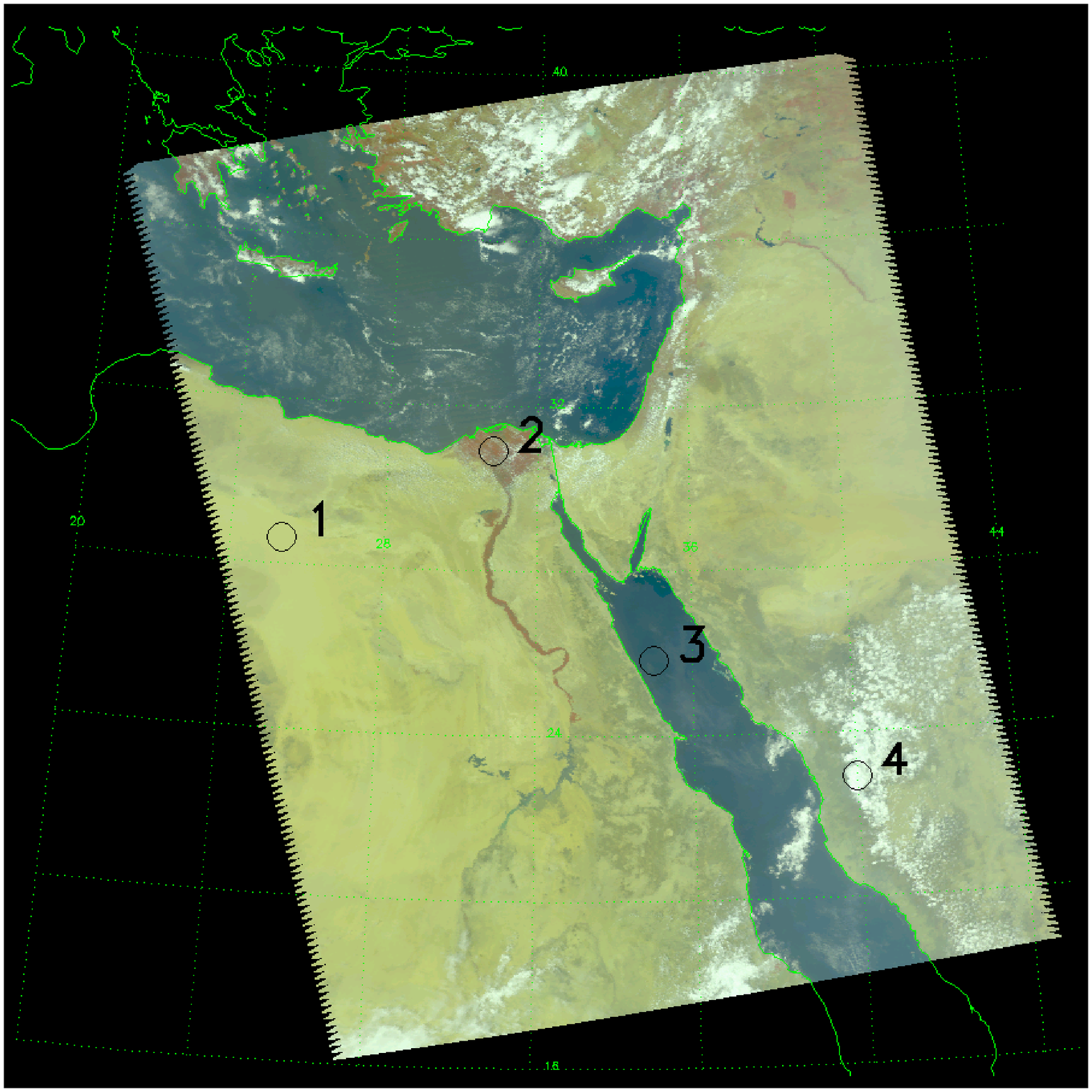}
\includegraphics[width=.6\textwidth]{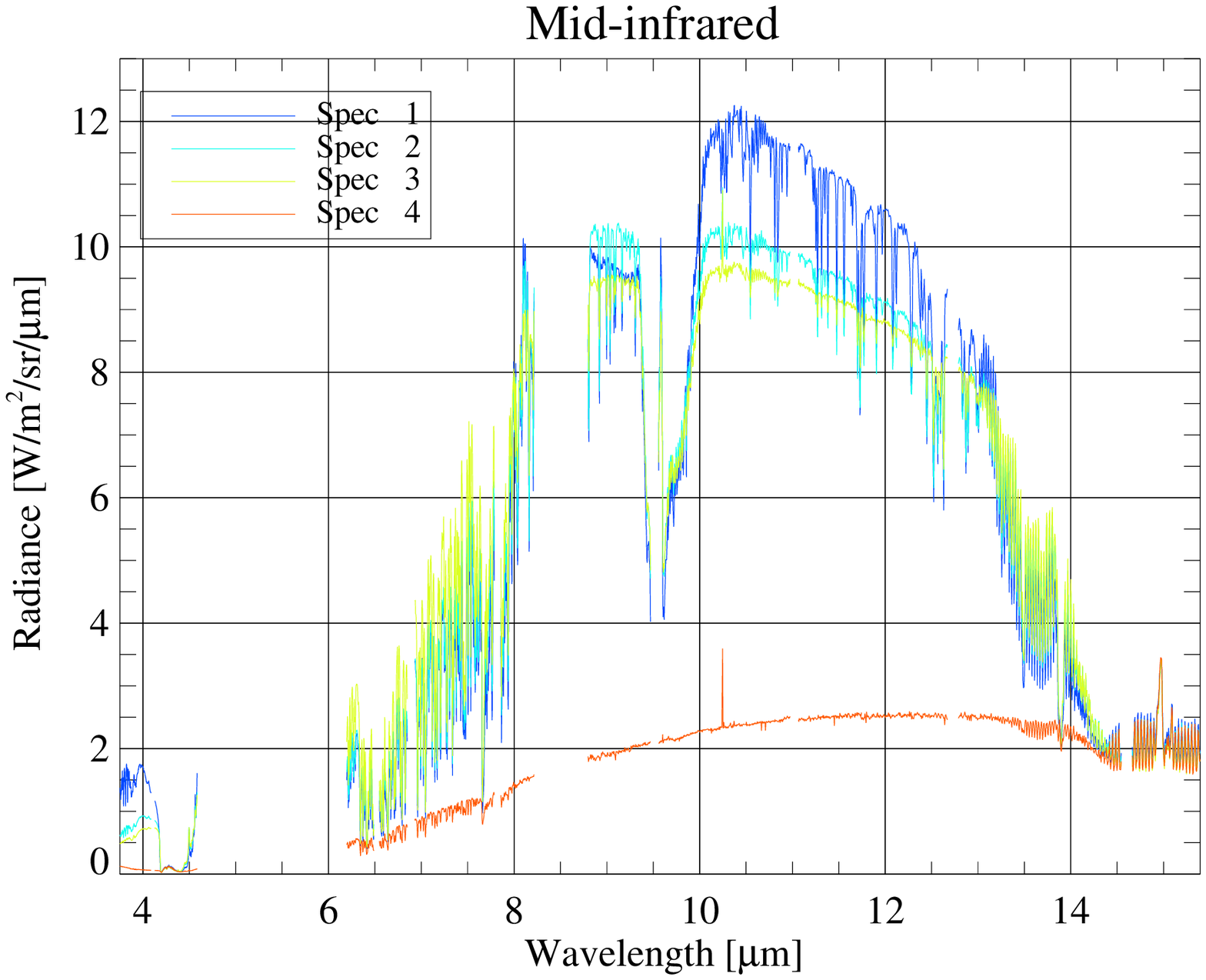}
\includegraphics[width=.6\textwidth]{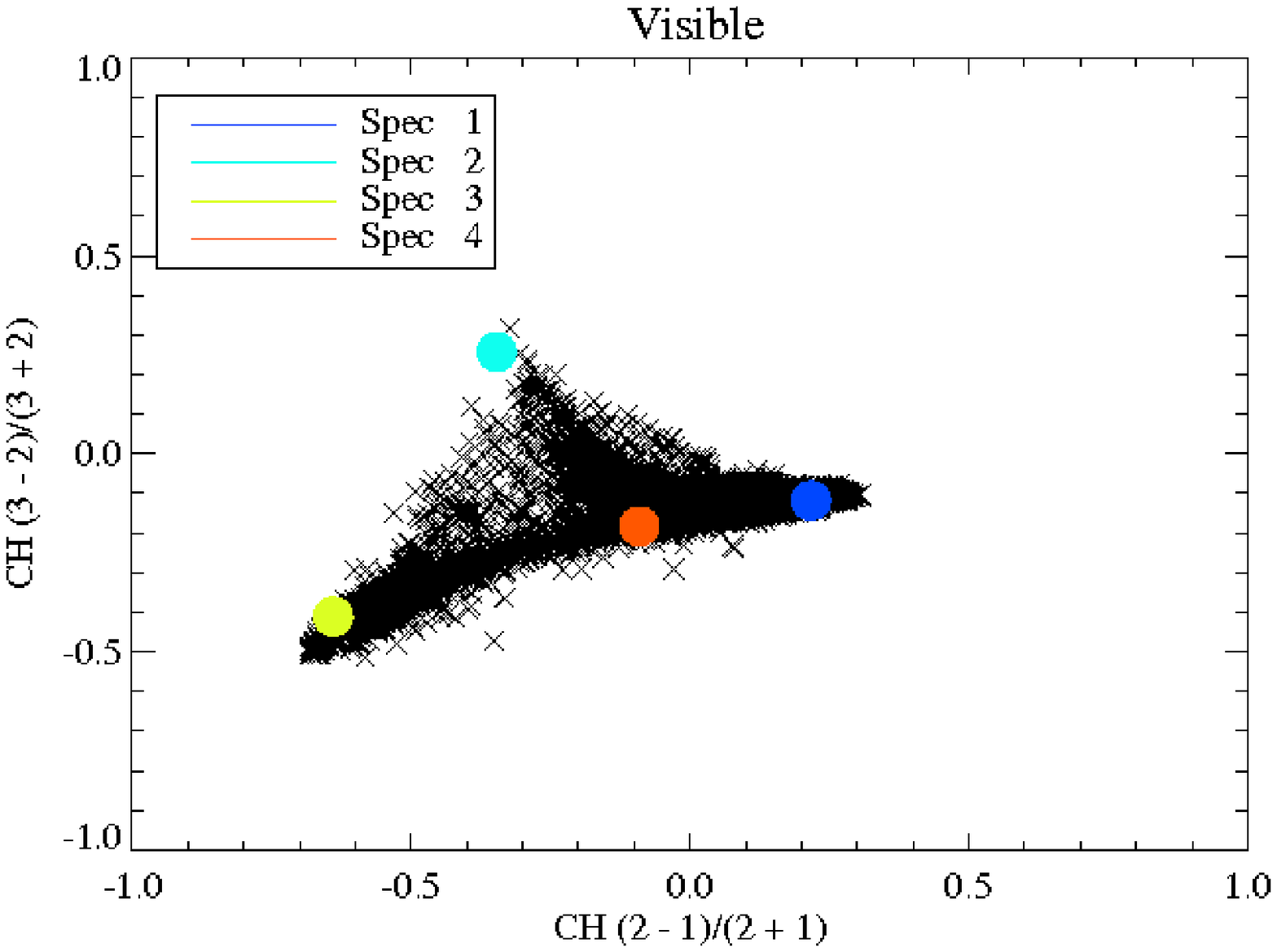}
\end{center}
\caption{The top panel shows a visible image of a region observed with AIRS.  The regions labeled 1-4 are the Sahara Desert, the Nile Delta, clear ocean, and cloud covered, respectively.  The middle panel shows the mid infrared spectra from the regions labeled in the visible map.  The bottom panel shows normalized visible/near-infrared channel differences (see Paper II) for all of the scenes in the figure with the 4 labeled regions indicated.}
\label{egypt}
\end{figure}

Although future terrestrial planet finding telescopes must be designed
so that they can characterize diverse
types of terrestrial planets, Figure~\ref{egypt} shows that a
relatively small region of
Earth can display a large diversity of spectral features
in the visible and infrared.
However, since the first generation of terrestrial planet characterization missions
will not spatially resolve the surface features on
the planets they observe, only one spectrum will be observed for the whole visible disk.  
This paper examines the observable mid-infrared spectroscopic properties
and Paper II (Hearty et al. 2008) examines the observable visible/near-infrared
photometric properties.

\clearpage
\thispagestyle{empty}
\setlength{\voffset}{-17mm}
\clearpage
\setlength{\voffset}{0mm}

Whole Earth mid-infrared spectra have been directly obtained
by the Galileo satellite (Sagan et al. 1993) and 
the Mars Global Surveyor satellite with the TES instrument \citep{chr97}.
However, the spectral resolution of these investigations
was not sufficient for detailed analysis and they only viewed
Earth from a limited set of viewing geometries and seasons.
``Earthshine'' spectra are able to observe seasonal and rotational variations
\citep{woo02,pal03,mon05}, however, they are limited to the visible near-IR
region of the spectrum and they can only view earth edge-on.
We have used directly
observed spectra of Earth obtained with the AIRS
to provide a detailed set of spectra to test
existing models \citep{for01,des02,tin05,tin06a,tin06b}.

Although there
is no reason to expect that we
will observe a planet exactly like Earth or at the same
evolutionary stage as Earth, 
the detailed spectra of Earth obtained by AIRS allow us to explore all of the
biomarkers that may be seen when TPF-like missions observe
an Earth-like planet orbiting another star.  Thus, our AIRS spectra can serve as
a library of sample spectra from a life-bearing habitable
planet.

Section~\ref{obs} describes the AIRS observations we used and how we converted the observed
radiances to whole Earth spectra.  We also compare a whole earth spectrum generated from AIRS
to a TES/MGS
observation of Earth for a similar season and viewing geometry from 1996. 
Section~\ref{spectra} describes the spectral signatures in the AIRS
whole Earth spectra and what signatures will be observable with future telescopes capable of
observing extrasolar terrestrial planets.
Sections~\ref{rotational} and~\ref{seasonal}
examine the rotational and seasonal variations that could be detected in
observations of an Earth-like extrasolar planet.  Section~\ref{conclusion} summarizes our findings.

\section{Observations and Data Reduction}
\label{obs}

We investigated one day of AIRS spectra from the 26th of each month from September 26,
2004--August 26, 2005 using version 4 of the AIRS data products which are available
from the Goddard Earth Sciences Data and Information Services
Center\footnote{http://disc.sci.gsfc.nasa.gov/AIRS/data\_products.shtml}.
We placed the AIRS radiances onto the HEALPix\footnote{http://healpix.jpl.nasa.gov} grid \citep{gor05} with Nside = 32
(12,288 pixels covering the earth).  
The HEALPix grid is an equal area grid that allows us to
simulate how the Earth would look from different
perspectives (e.g., edge-on and pole-on views) and easily calculate disk averaged spectra.
Higher spatial resolution
grids resulted in negligible differences in the disk averaged spectra.
We produced grids to simulate (1) a ``cloud-free'' Earth and
(2) a ``fully cloud-covered'' Earth using only clear or cloudy spectra culled from the full data set,
and (3) a ``normal'' Earth using all of the data.
We selected the clear and cloudy spectra by examining the
AIRS ``Level 2'' data that provides an effective cloud fraction ({\bf CldFrcStd}) for each scene.
We defined ``cloud-free'' observations as those with a cloud fraction $<$
0.05 and ``fully cloud-covered'' observations as those with a cloud fraction $>$ 0.95. 
Because of the paucity of observations that satisfy the clear (and cloudy) cases we used
4 days of AIRS observations to
fill-in the grid before interpolating over the missing data.

\begin{figure}
\begin{center}
\includegraphics[width=.45\textwidth,angle=90]{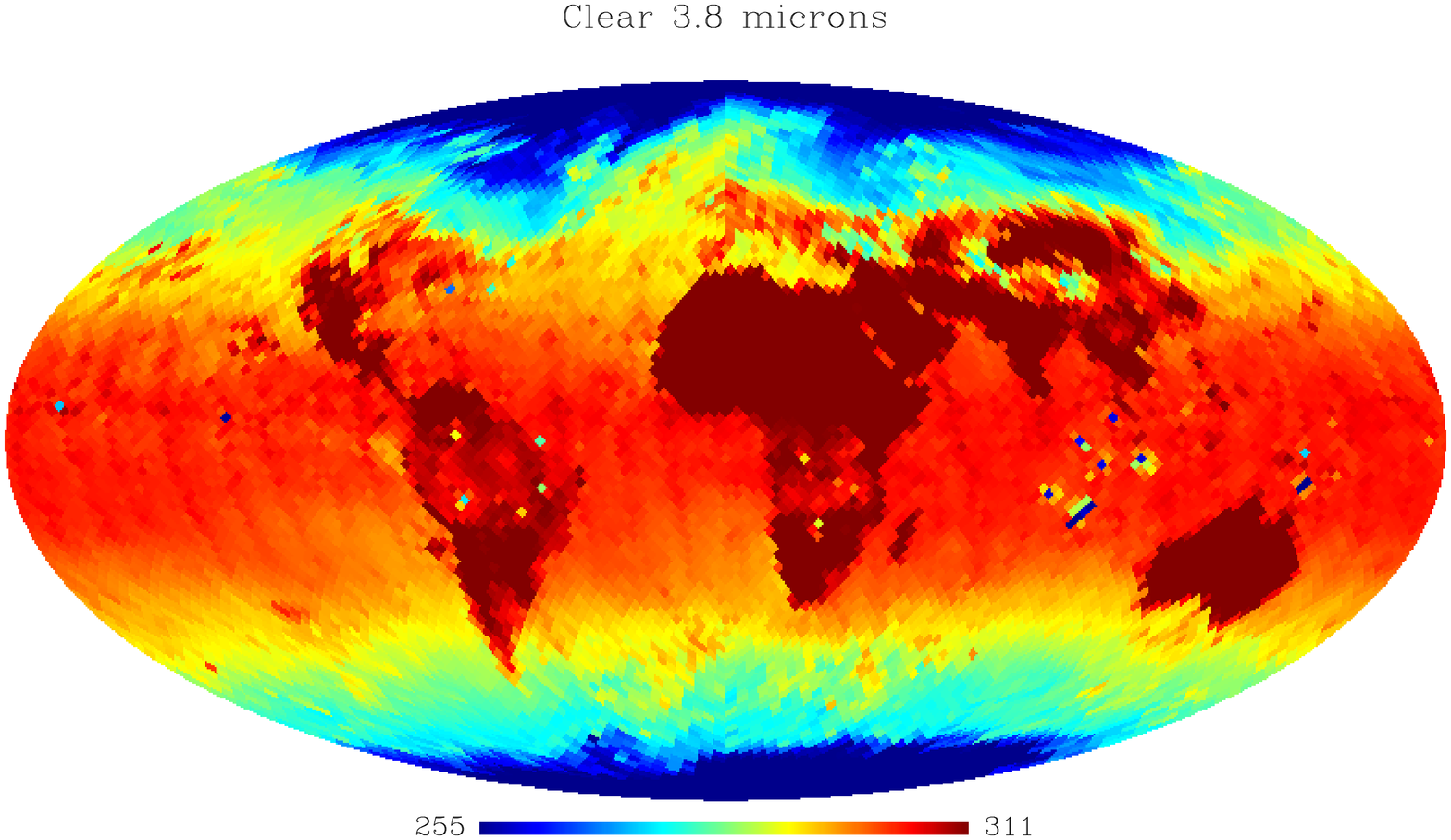}
\includegraphics[width=.45\textwidth,angle=90]{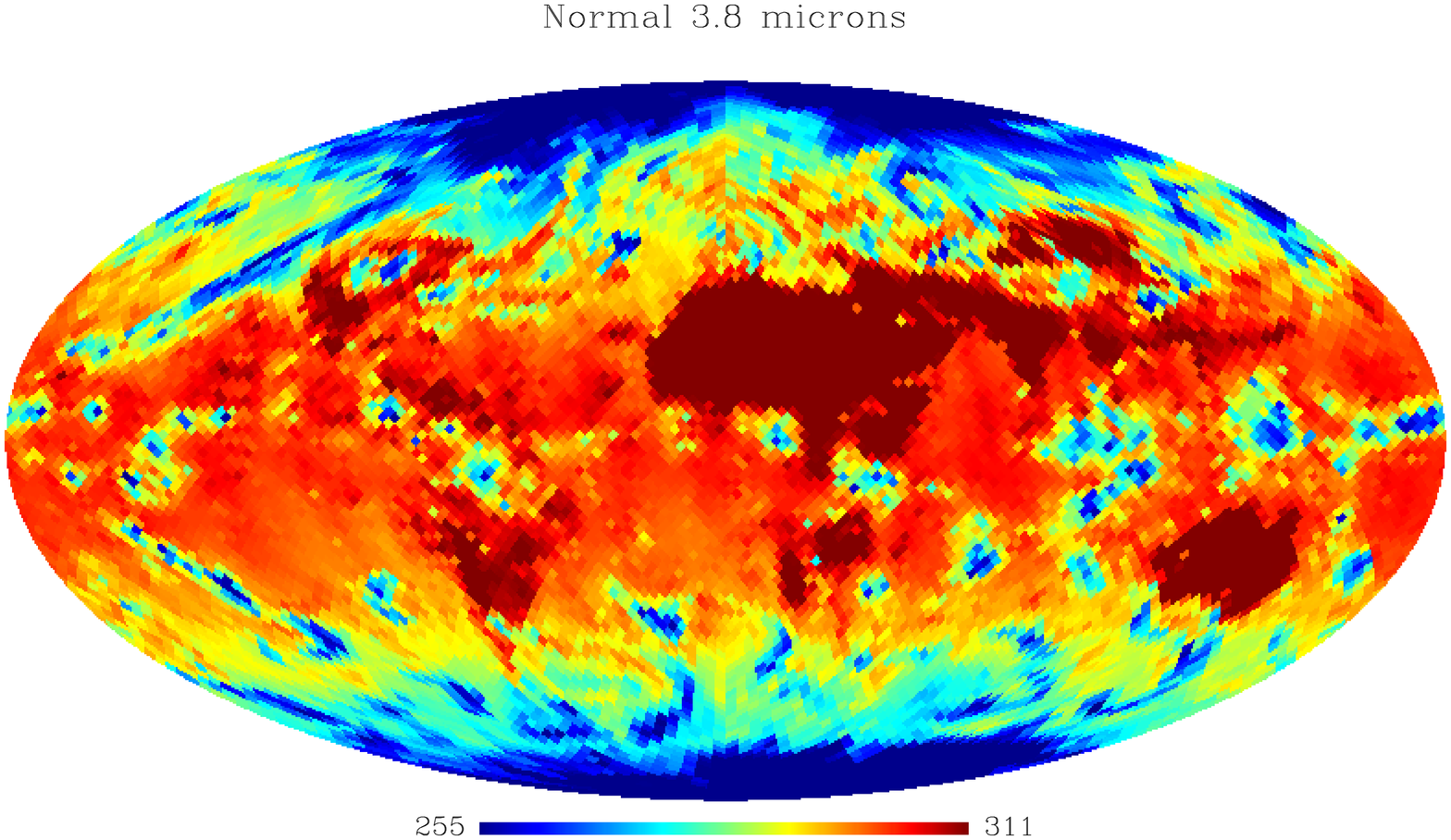}
\includegraphics[width=.45\textwidth,angle=90]{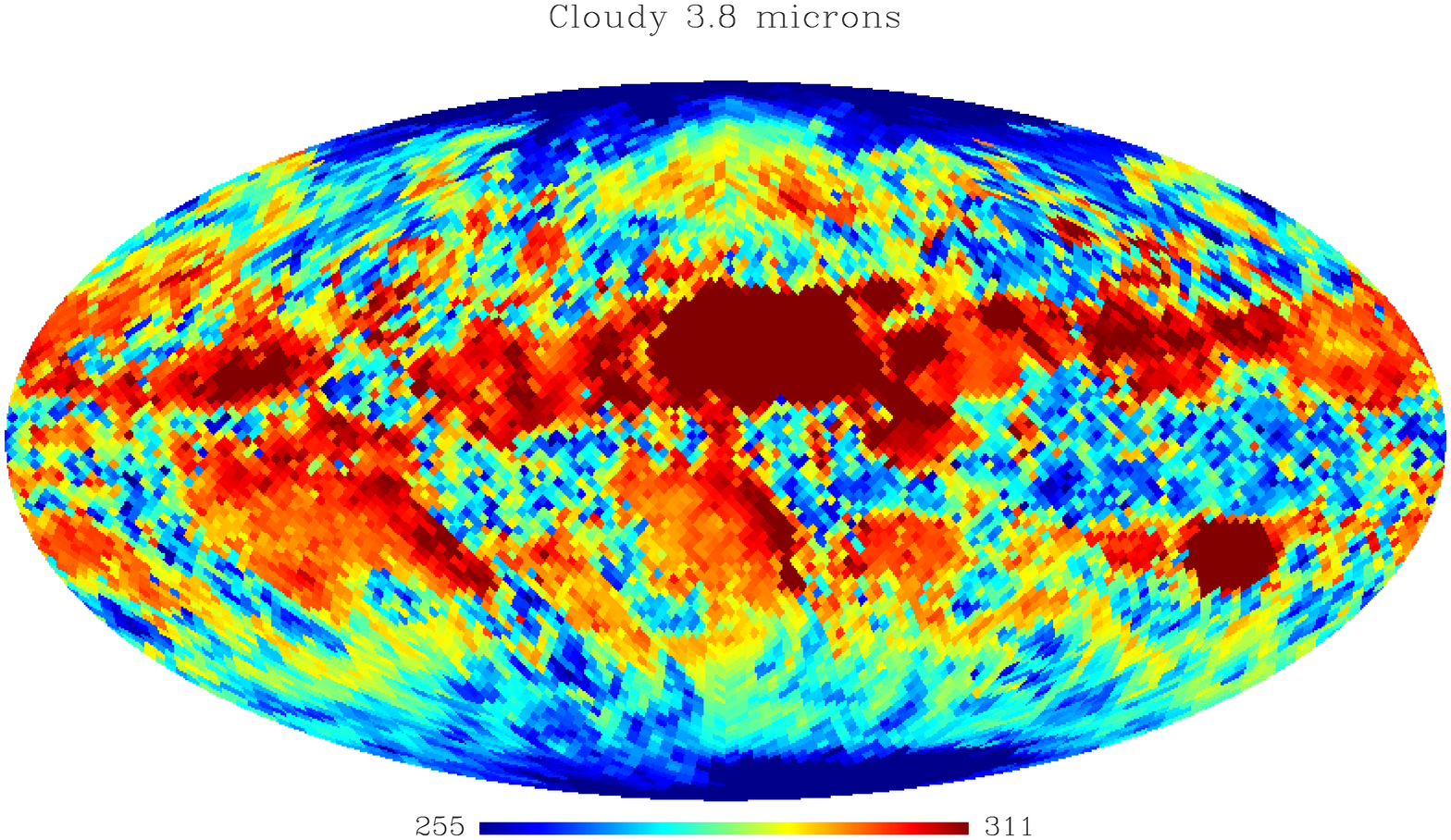}
\end{center}
\caption{Maps of the daytime infrared brightness temperature [Kelvin] of the Earth at 3.8 $\mu$m (2616 cm$^{-1}$) are displayed for
clear scenes (top), the actual earth (middle), and a cloud covered earth (bottom).} 
\label{ir_day}
\end{figure}

Figure~\ref{ir_day} shows global maps of the gridded data for
the clear, normal, and cloudy cases.  The clouds tend to mask surface features in the
infrared brightness temperature maps
of earth.  Because Version 4.0.9.0 of the AIRS processing software sometimes misidentifies
clear scenes as cloudy over very warm land, the significance of the masking of surface features
is underestimated for some regions (e.g., the Sahara and Australia).

\clearpage
\thispagestyle{empty}
\setlength{\voffset}{-12mm}
\clearpage
\setlength{\voffset}{0mm}

Since the AIRS data can be affected by cosmic rays, occasional noise events,
and the South Atlantic Anomaly, we did not include any radiances with
questionable calibration quality
assessment parameters.  Specifically,
we only included radiances for which the {\bf CalFlag} parameter is 0.

\begin{figure}
\includegraphics[width=.9\textwidth]{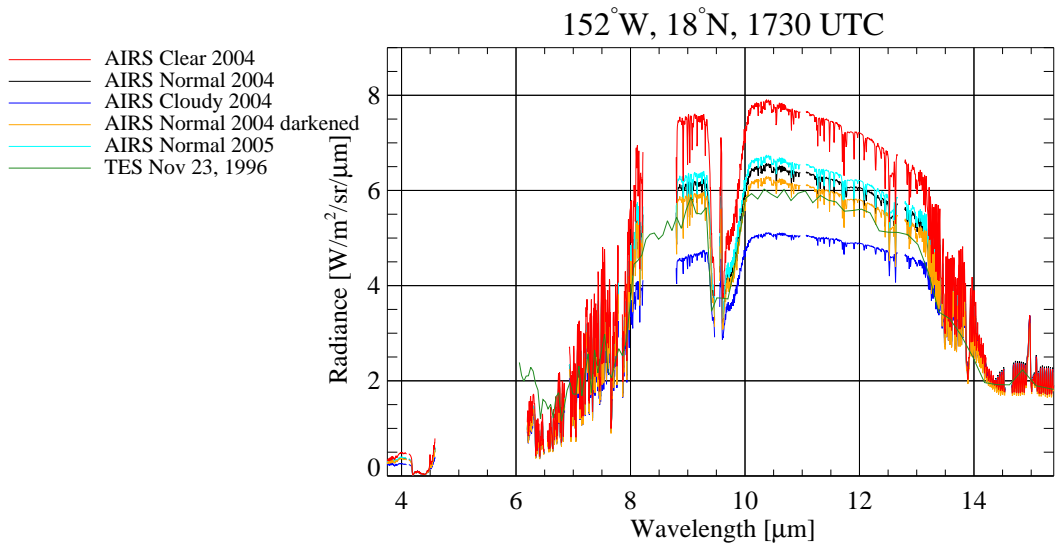}
\caption{AIRS observations from November 26, 2004 and 2005 are compared with a TES observation of Earth from November 23, 1996.}
\label{tescomparison}
\end{figure}

Figure~\ref{tescomparison} shows a comparison with a spectrum of Earth obtained with the TES instrument
on the Mars global surveyor to AIRS whole earth spectra constructed for a similar viewing geometry
for clear, normal, and cloudy cases from November 2004 and a normal cloud cover case
from November 2005.  The figure also includes
a disc averaged spectrum calculated using a simple limb darkening parameterization
adapted from \citet{hod00} where the limb adjusted radiance R($\theta$) is calculated from the
radiance at nadir R(0) as follows:
$$ R(\theta) = \lambda(\theta)  \times R(0), $$
where
$$ \lambda(\theta) = 1 + 0.09 \times ln[cos(\theta)] $$
and $\theta$ is the zenith angle.  

The observed differences between the AIRS and TES spectra could be due to slight differences in the assumed
viewing geometry, cloud cover, limb effects, or calibration.  
Since there are differences between the AIRS
spectra from November 2004 and November 2005 which have identical viewing geometries and calibration, we know that
year-to-year spectral variations can be discerned.
Although the simple limb darkening model seems to bring the AIRS and TES spectra into better agreement
the observations made by the two instruments are separated by more than 10 years,
thus the differences could also be due to changes in the cloud
cover which can also significantly affect the spectra.  A more accurate treatment of the limb
effects is beyond the scope of this study.  Moreover, since it is likely to be a small correction (relative
to the effects of clouds), we neglect limb effects for the rest of this paper.

\clearpage
\clearpage

\section{Spectral Signatures}
\label{spectra}

\begin{figure}
\includegraphics[width=.9\textwidth]{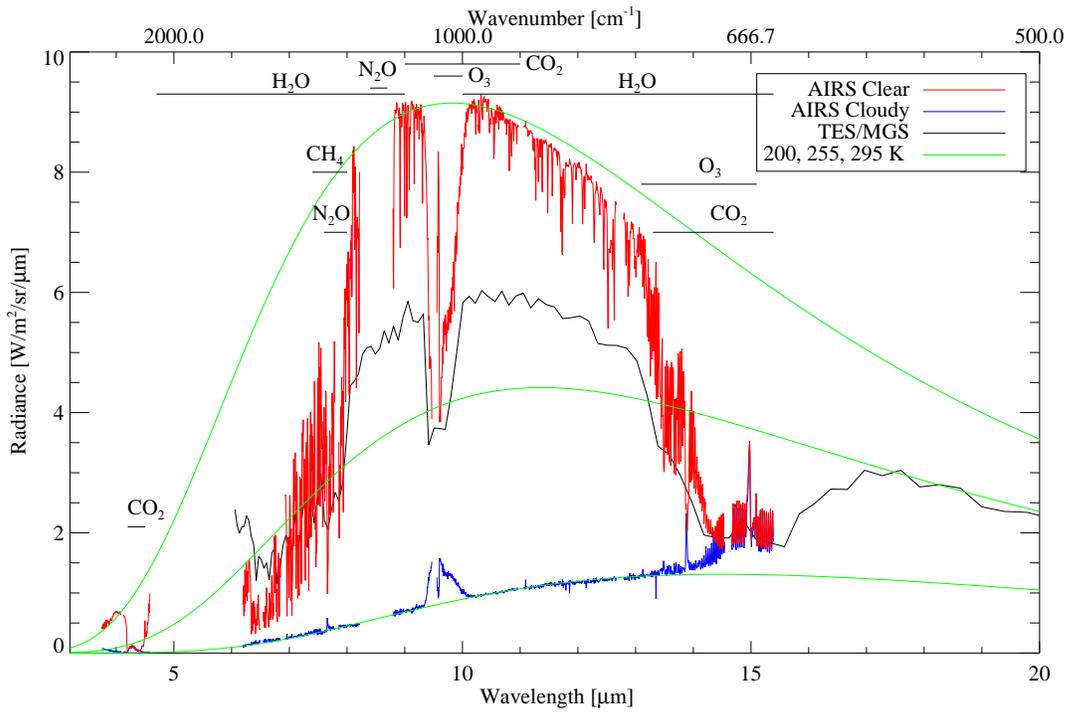}
\caption{Radiance spectra of two $\sim$ 15 km views of Earth obtained with AIRS
are displayed with a disk averaged spectrum of Earth obtained with TES/MGS and
three black body curves.
Black body fitting can give the surface temperature but the spectral lines may
be in absorption, emission, or not present.}
\label{AIRSbroad}
\end{figure}

\begin{figure}
\includegraphics[width=.49\textwidth]{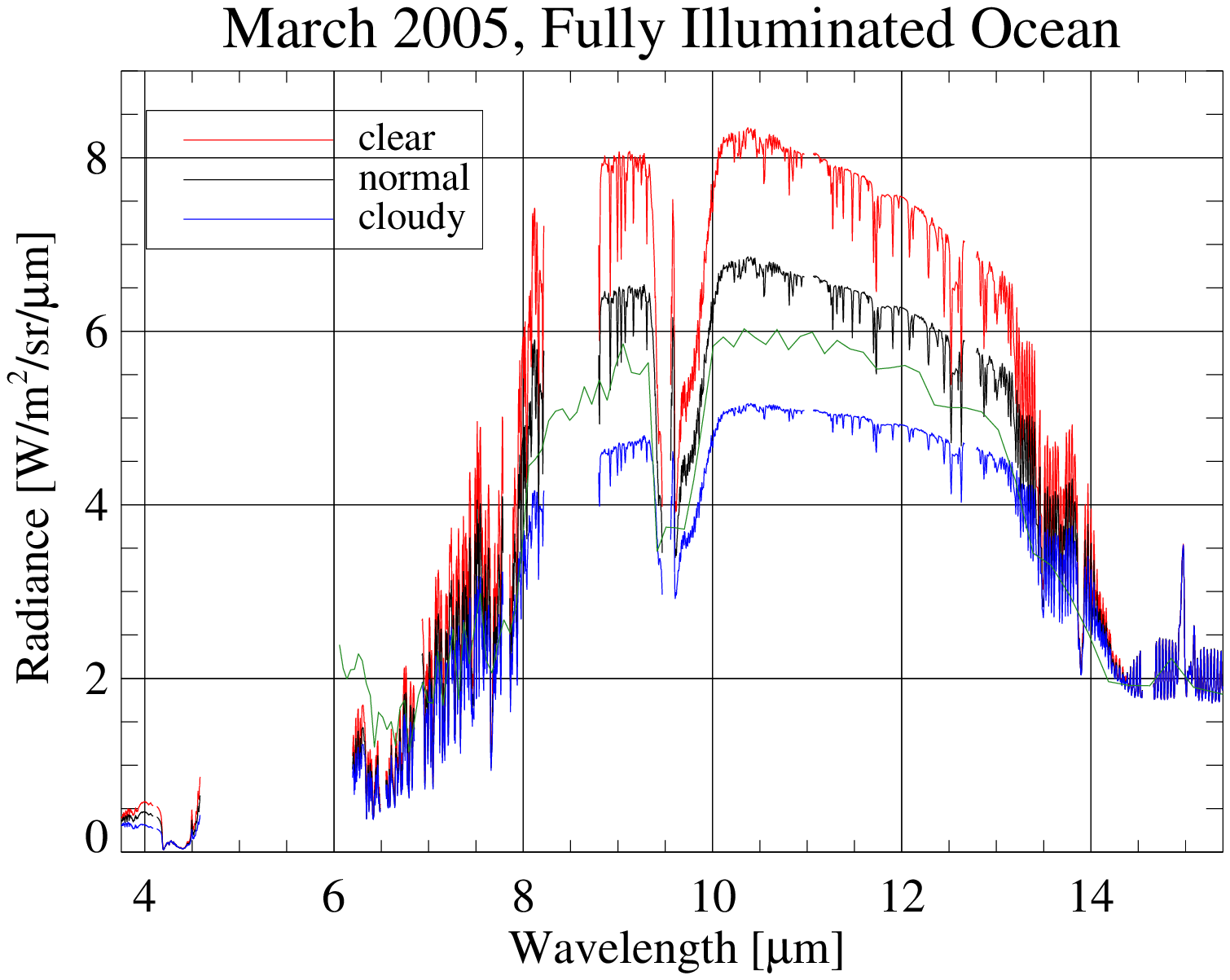}
\includegraphics[width=.49\textwidth]{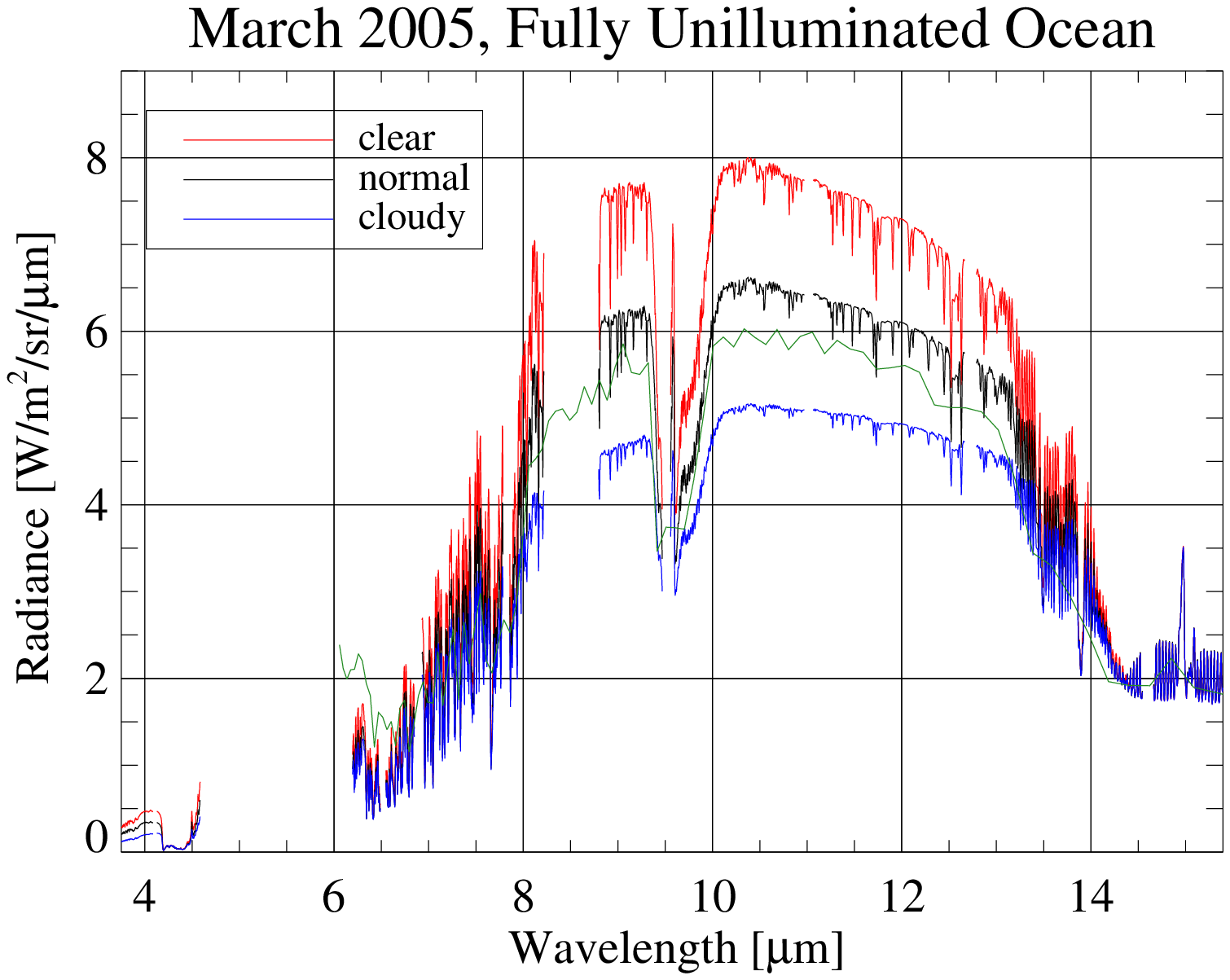} \\
\includegraphics[width=.49\textwidth]{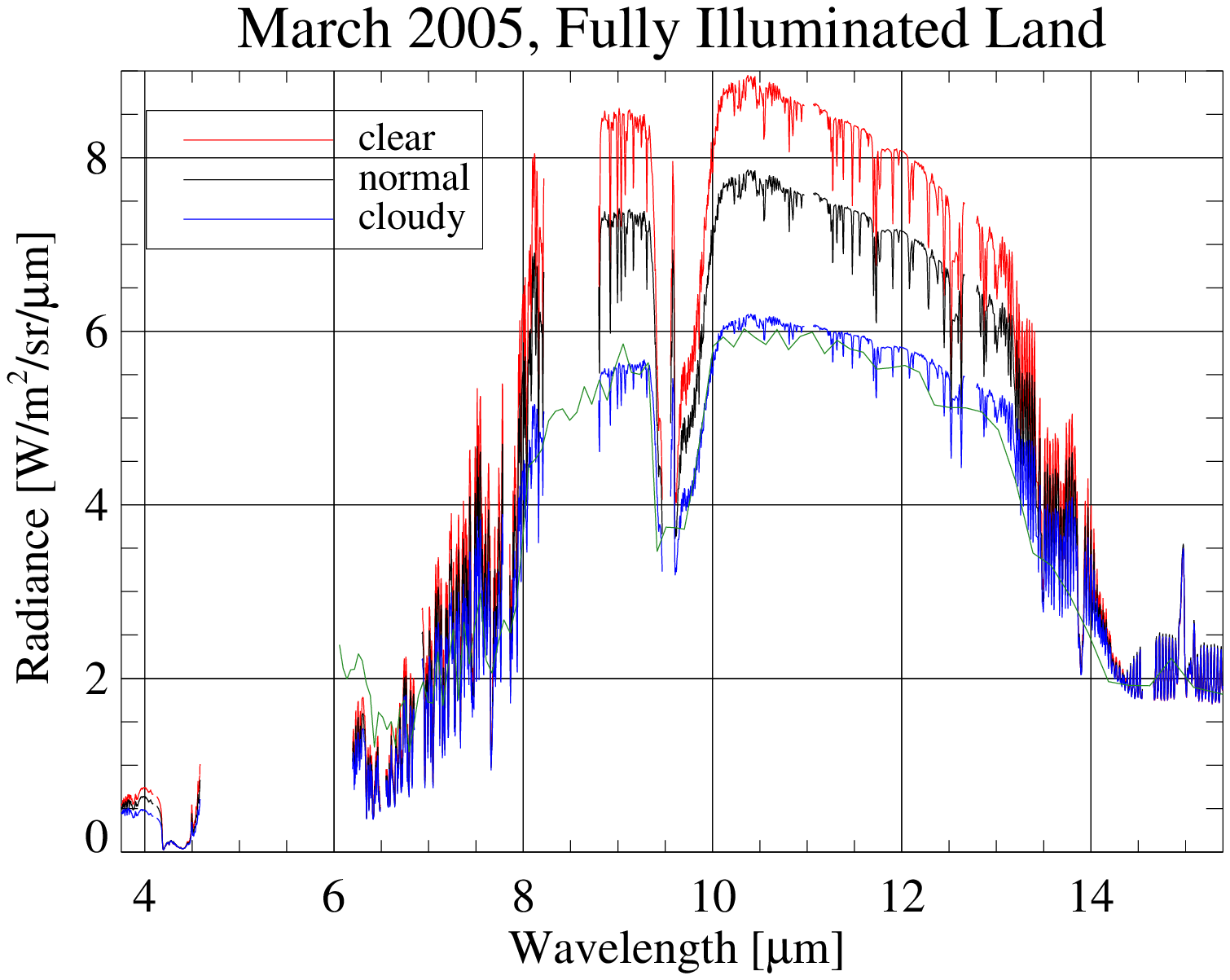}
\includegraphics[width=.49\textwidth]{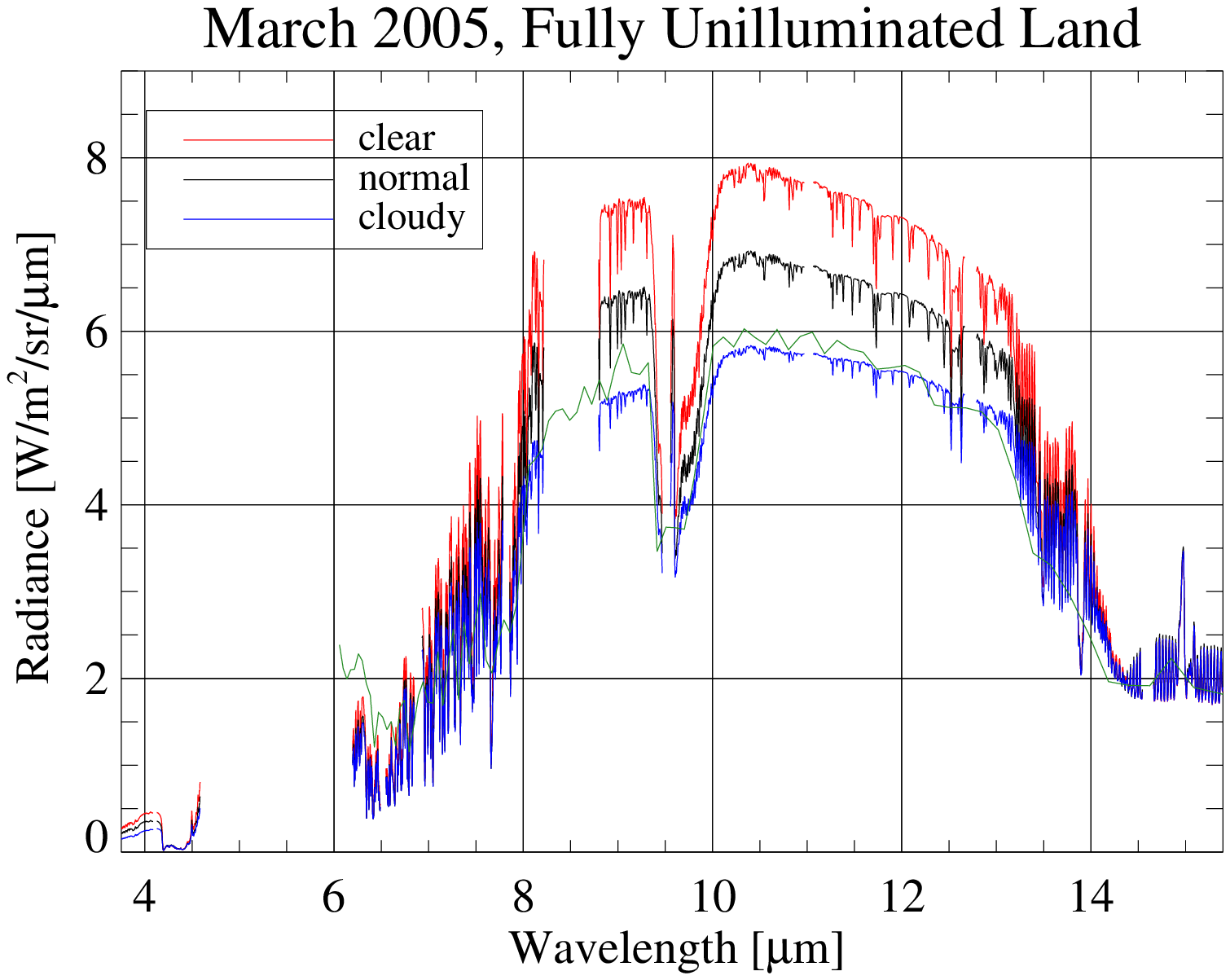}
\caption{Clear, Normal, and cloudy whole earth spectra are displayed
for 4 views of Earth on March 25, 2005:  Fully Illuminated Ocean (top left),
Fully Unilluminated Ocean (top right), Fully Illuminated Land (bottom left), and
Fully Unilluminated Land (bottom right).  The spectral variation due to clouds
is larger than the differences between day, night, land, or ocean.
The projected land fraction for the ``ocean'' case is $\sim$ 10\% and the projected
land fraction for the ``land'' case is $\sim$ 40\%}
\label{march_spec}
\end{figure}

The mid-infrared
observations with AIRS allow us to observe spectral signatures of
habitability and life (e.g., H$_2$O, CO$_2$, CH$_4$, and an apparent
temperature from the spectral shape).  
We used the 2004 HITRAN molecular transition line database to identify
the spectral features from the AIRS data (Rothman et al. 2005).
Figure~\ref{AIRSbroad} shows that
for clear and cloudy scenes the O$_3$ line, an important sign of life
can appear either in emission or absorption.
Since the Earth contains
regions that are clear and covered by dense clouds, the disk averaged AIRS spectra
enable us to determine the limitations in
characterizing the temperature of Earth-like planets with clouds.
This effect can be significant since the variation in the mid-infrared due to clouds
is greater than the differences between day and night.
Figure~\ref{march_spec} shows disk averaged broad-band mid-infrared spectra for fully illuminated
and fully unilluminated views of earth on March 26, 2005 for clear,
normal, and cloud covered cases.  The ``ocean view'' is centered above the pacific with
$\sim$ 10\% of the projected area filled with land and the
``land view'' is centered above Africa with $\sim$ 40\% of the projected area filled with land.
The day night difference is larger over land than over ocean.
We can also see from this figure that for an edge on view at opposition we will observe
a larger radiance variation for a rotating planet with an uneven distribution of
oceans and land (like Earth) than we will observe at conjunction.

\begin{figure}
\includegraphics[width=.32\textwidth]{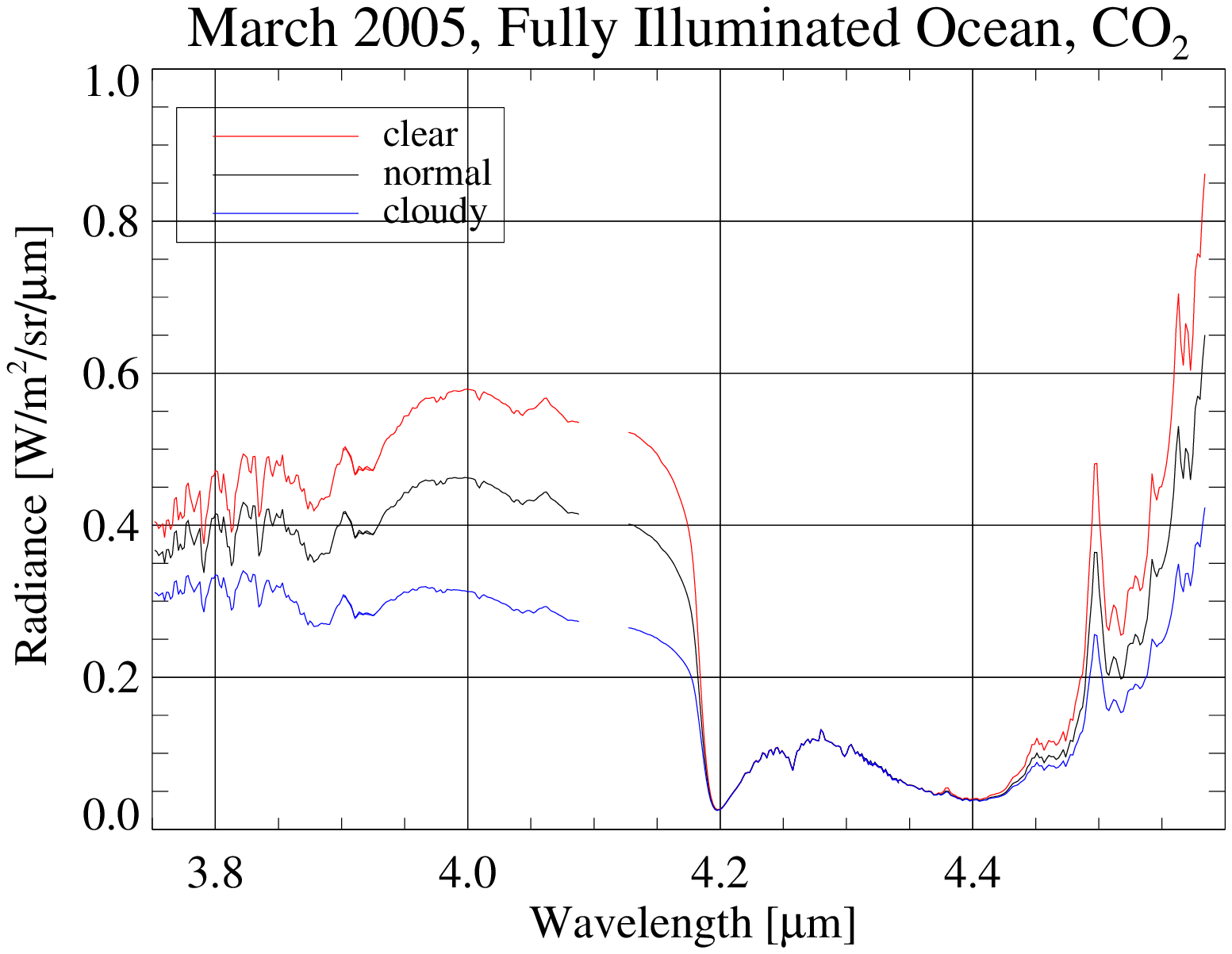}
\includegraphics[width=.32\textwidth]{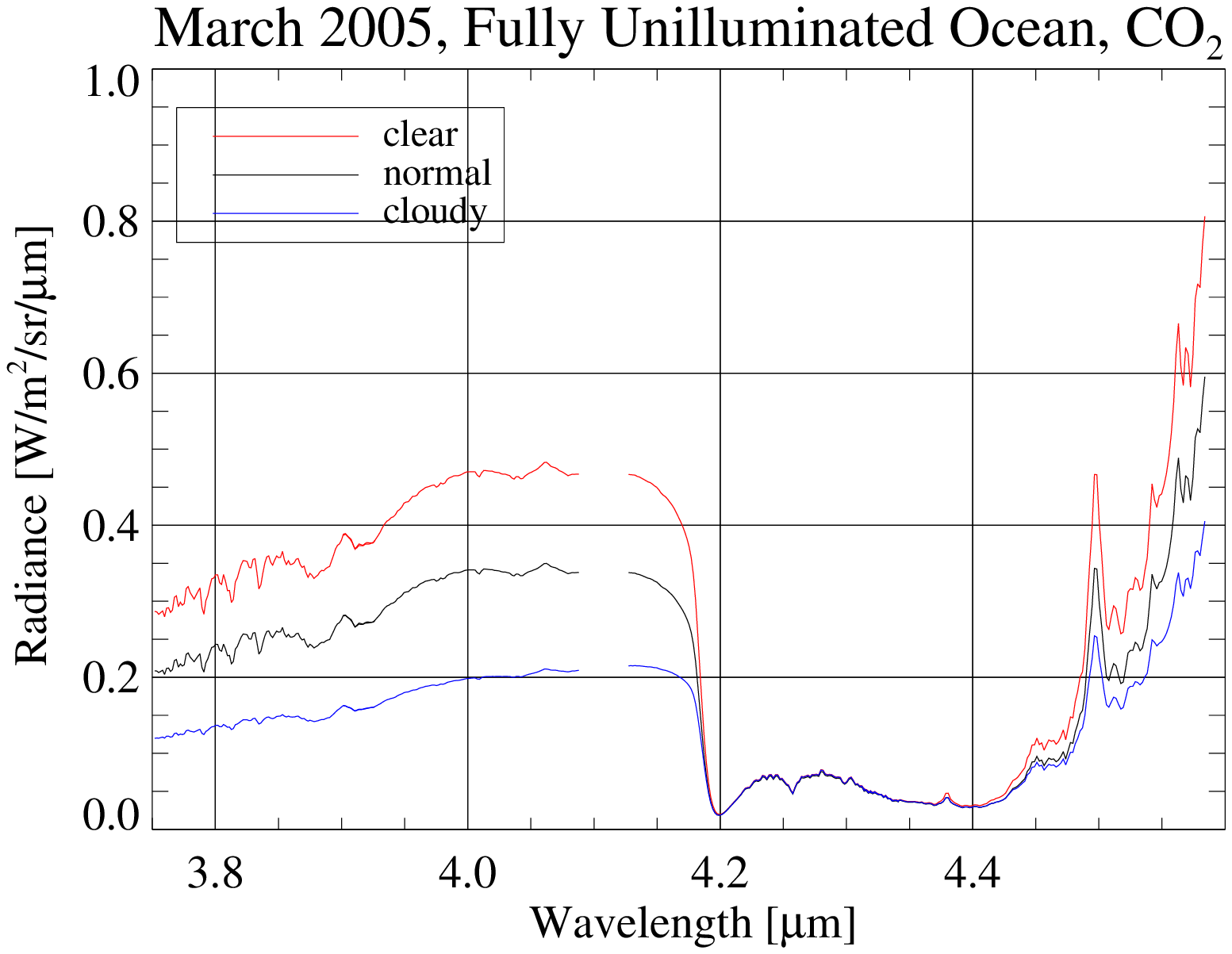}
\includegraphics[width=.32\textwidth]{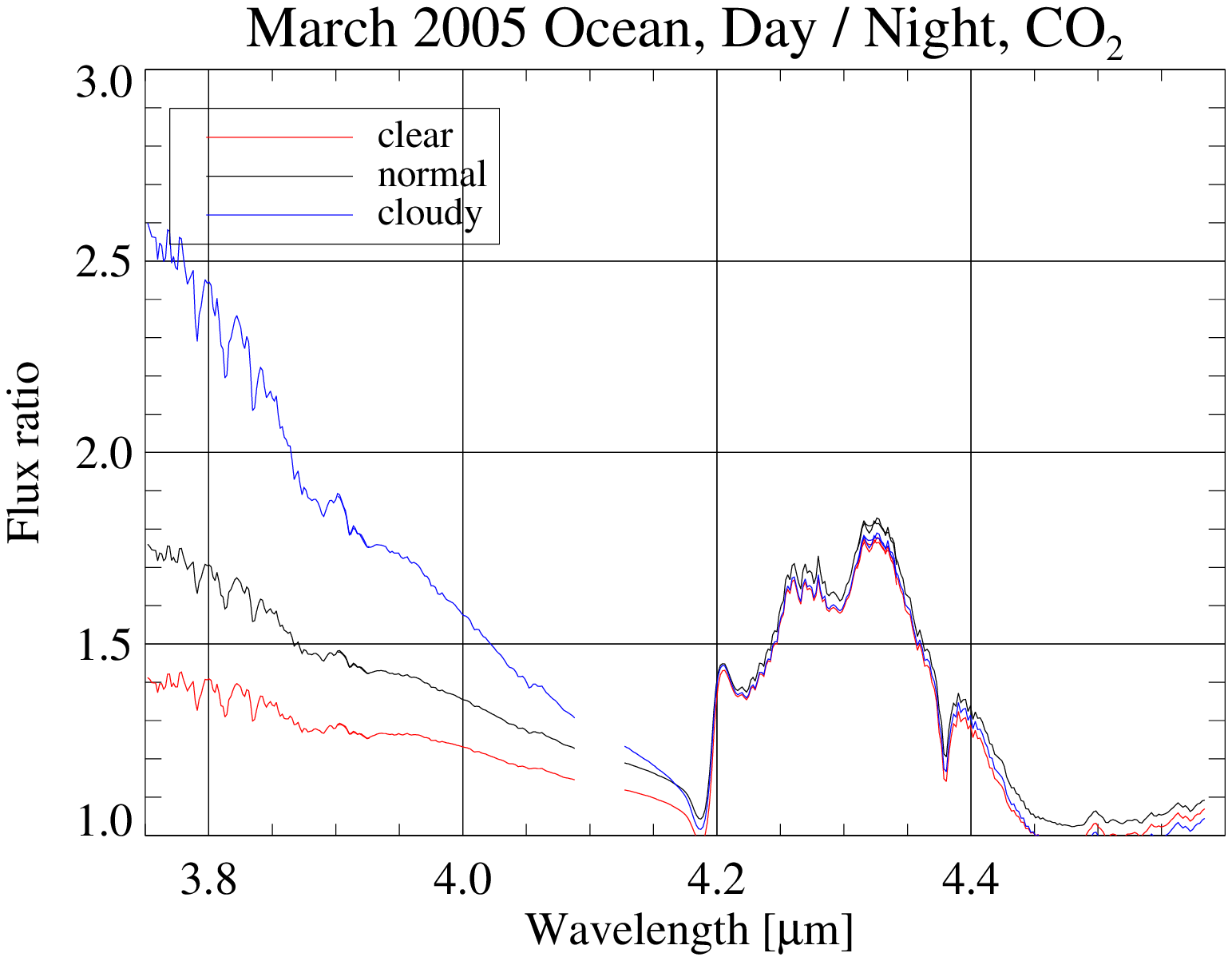} \\
\includegraphics[width=.32\textwidth]{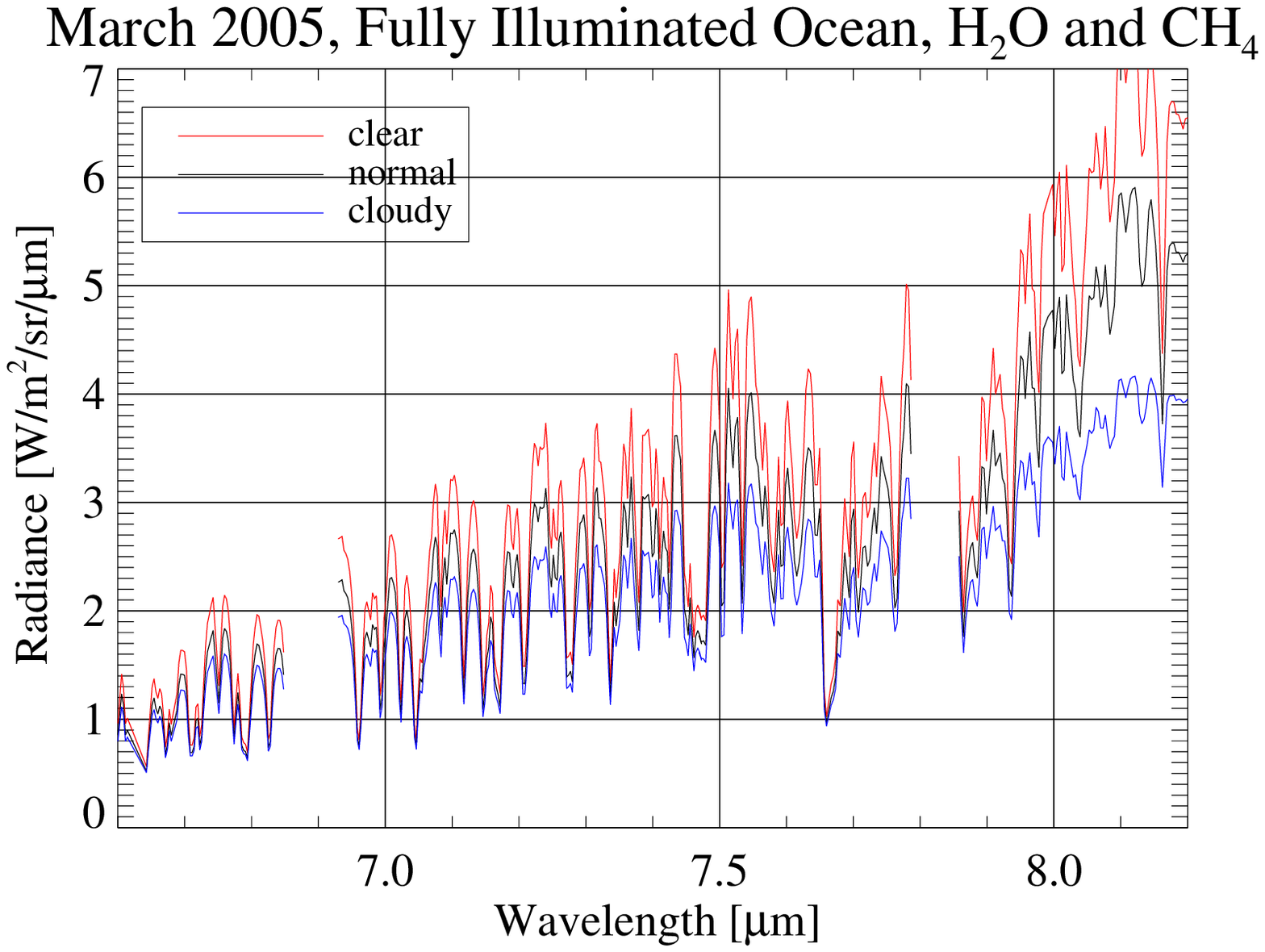}
\includegraphics[width=.32\textwidth]{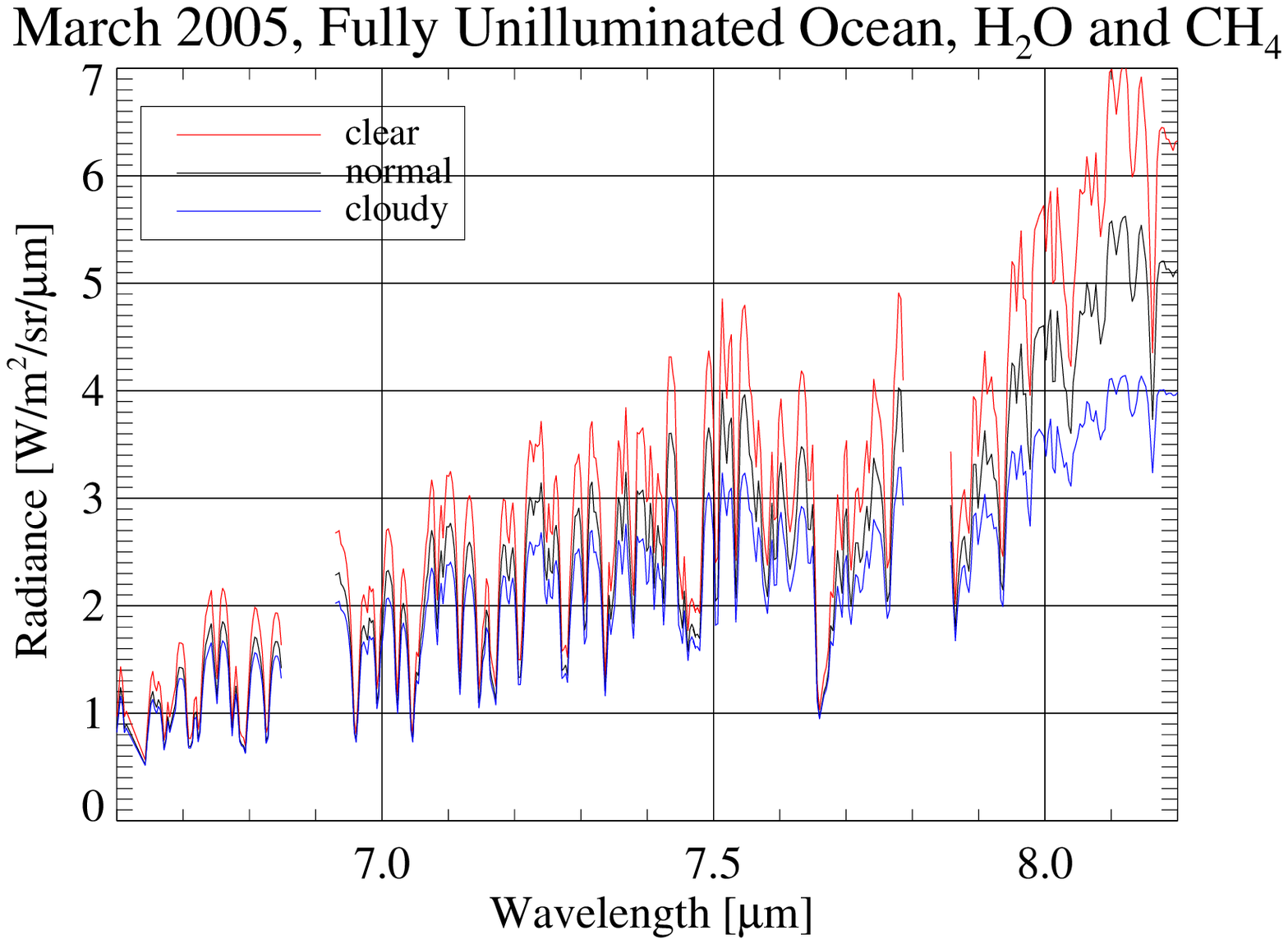}
\includegraphics[width=.32\textwidth]{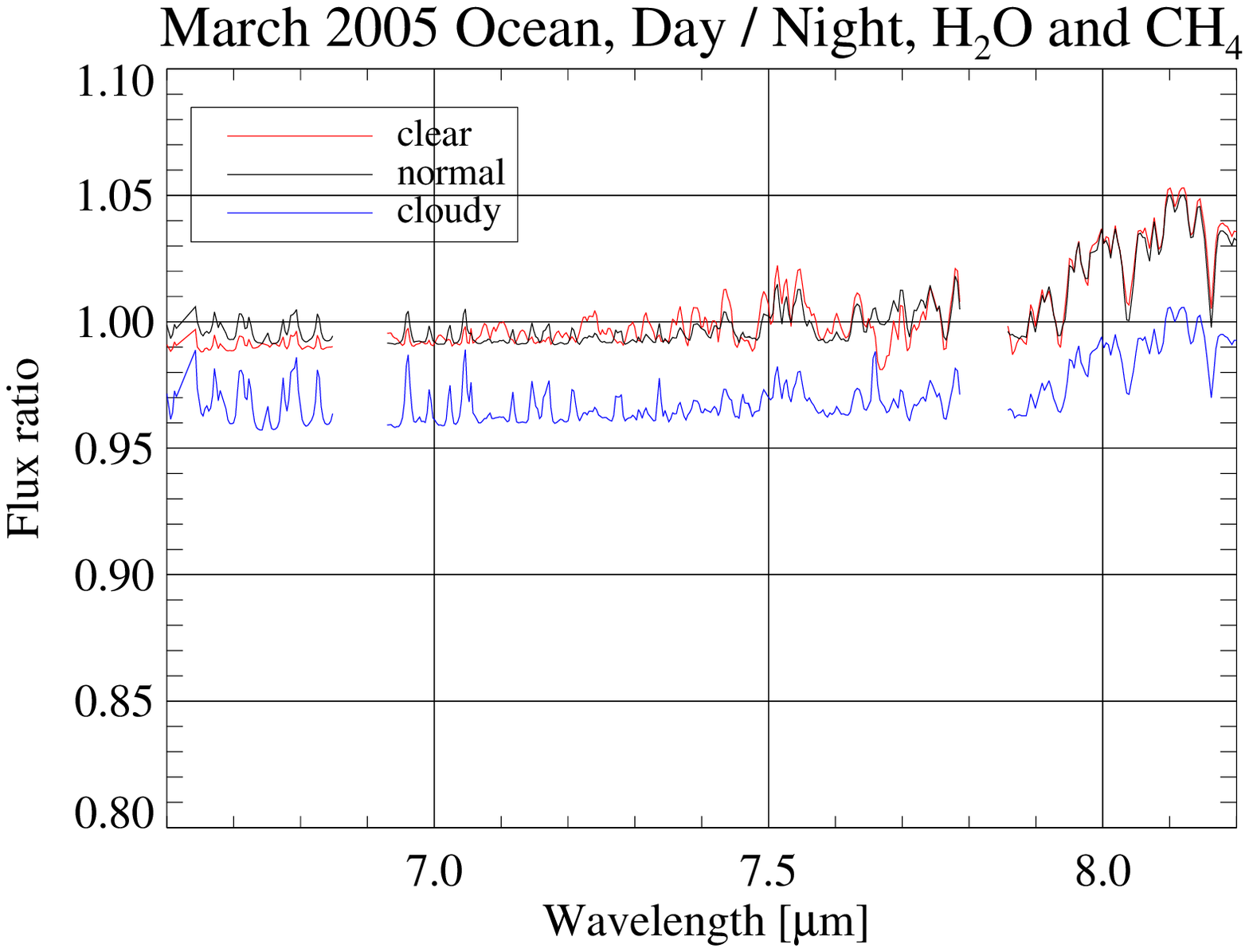} \\
\includegraphics[width=.32\textwidth]{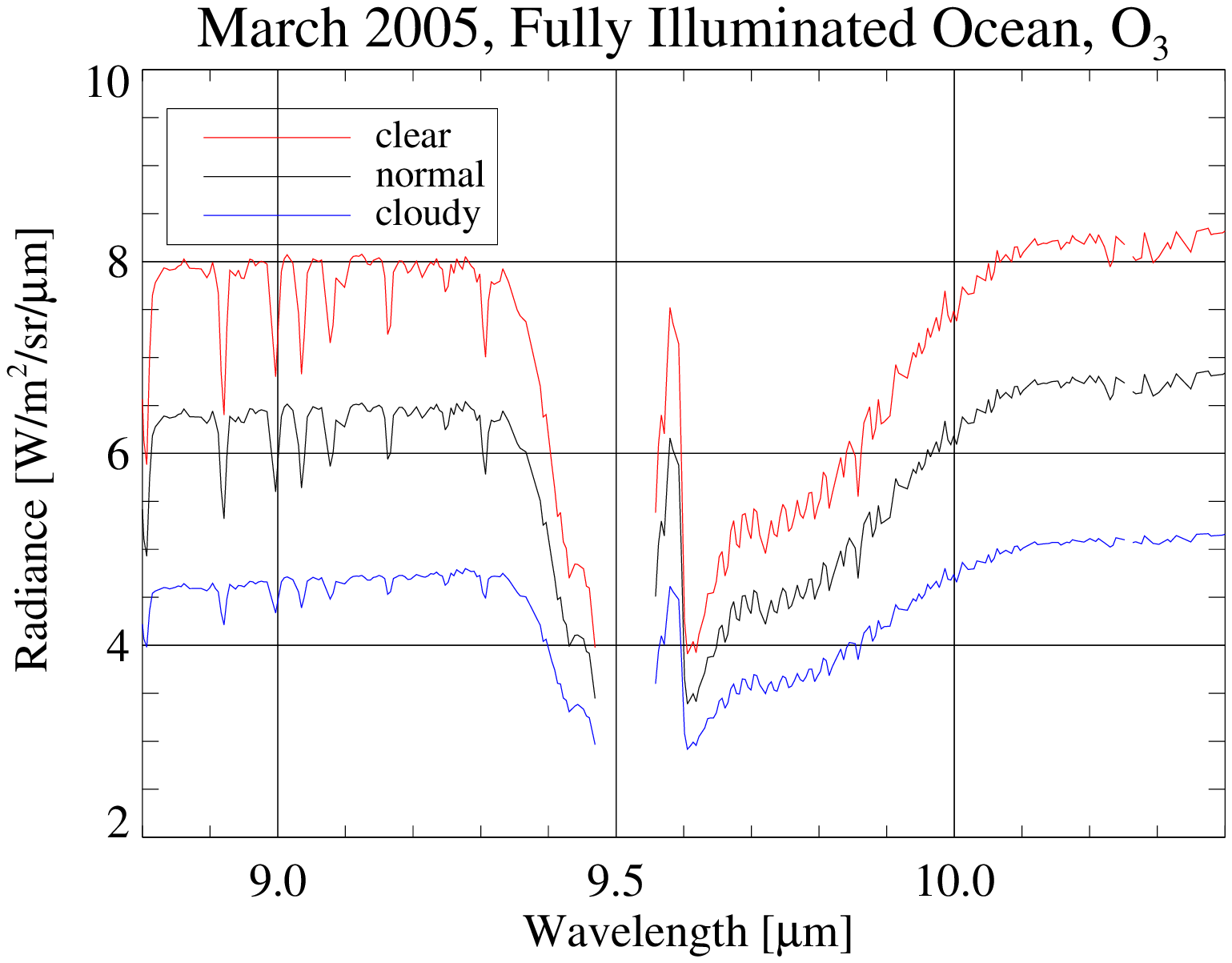}
\includegraphics[width=.32\textwidth]{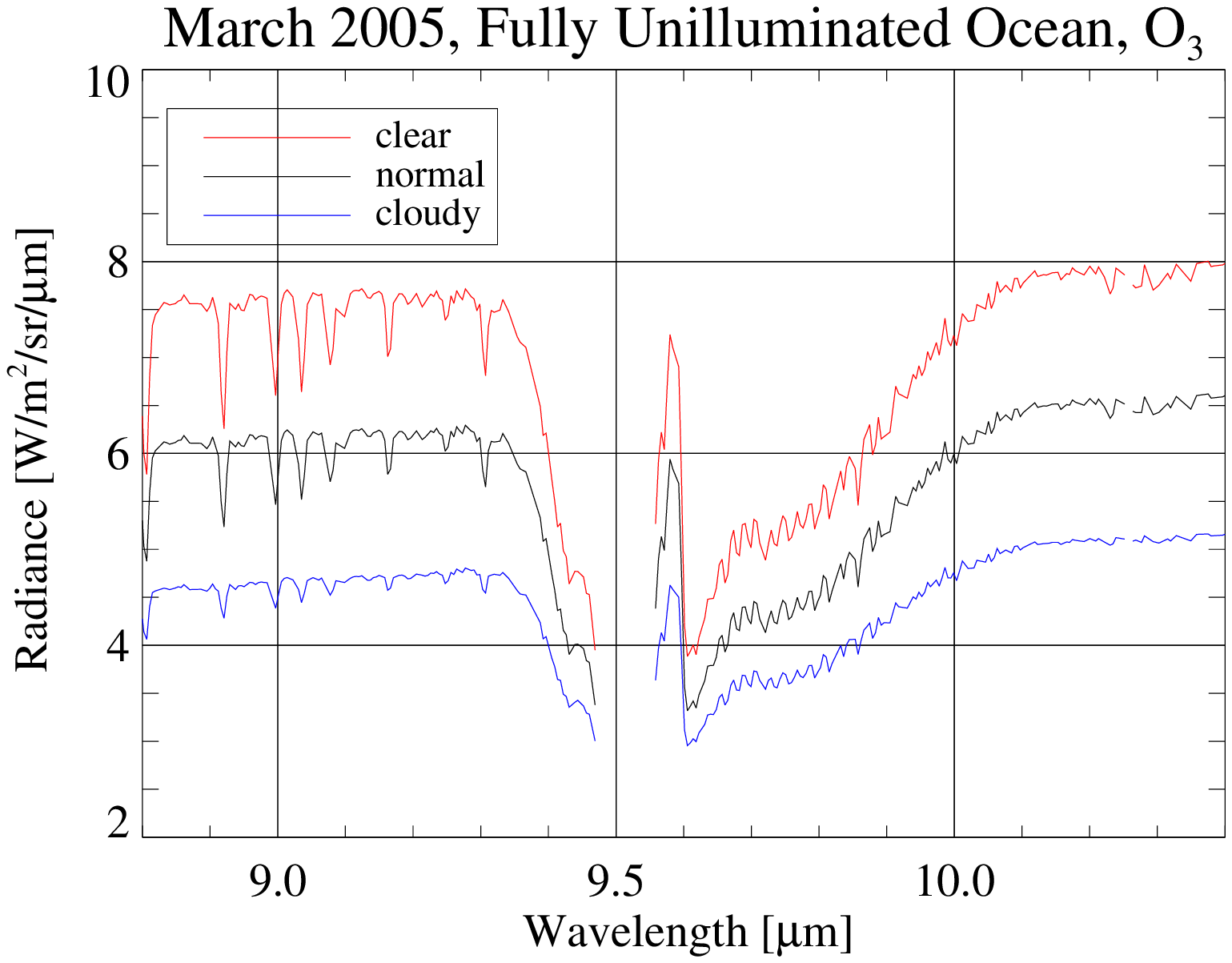}
\includegraphics[width=.32\textwidth]{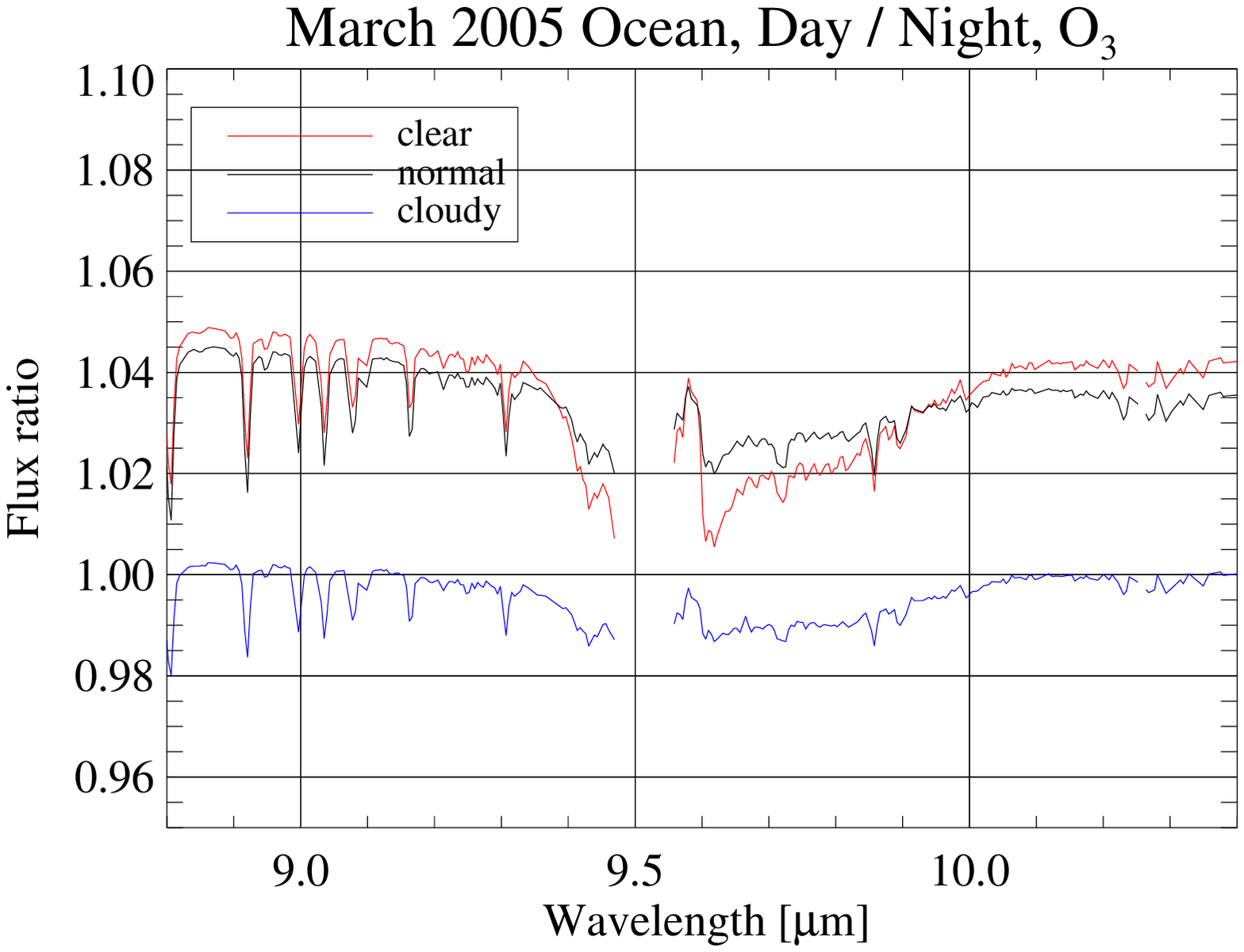} \\
\includegraphics[width=.32\textwidth]{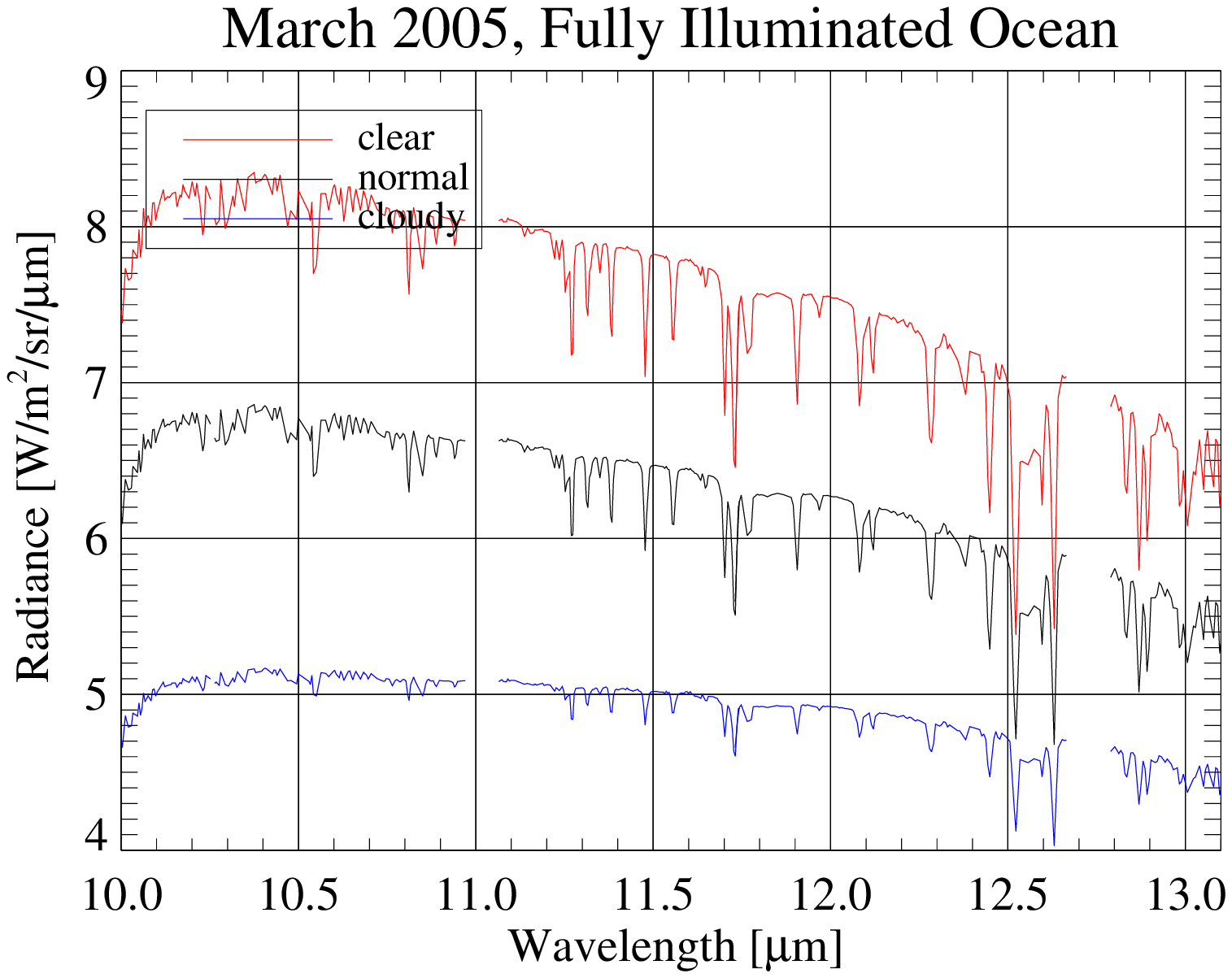}
\includegraphics[width=.32\textwidth]{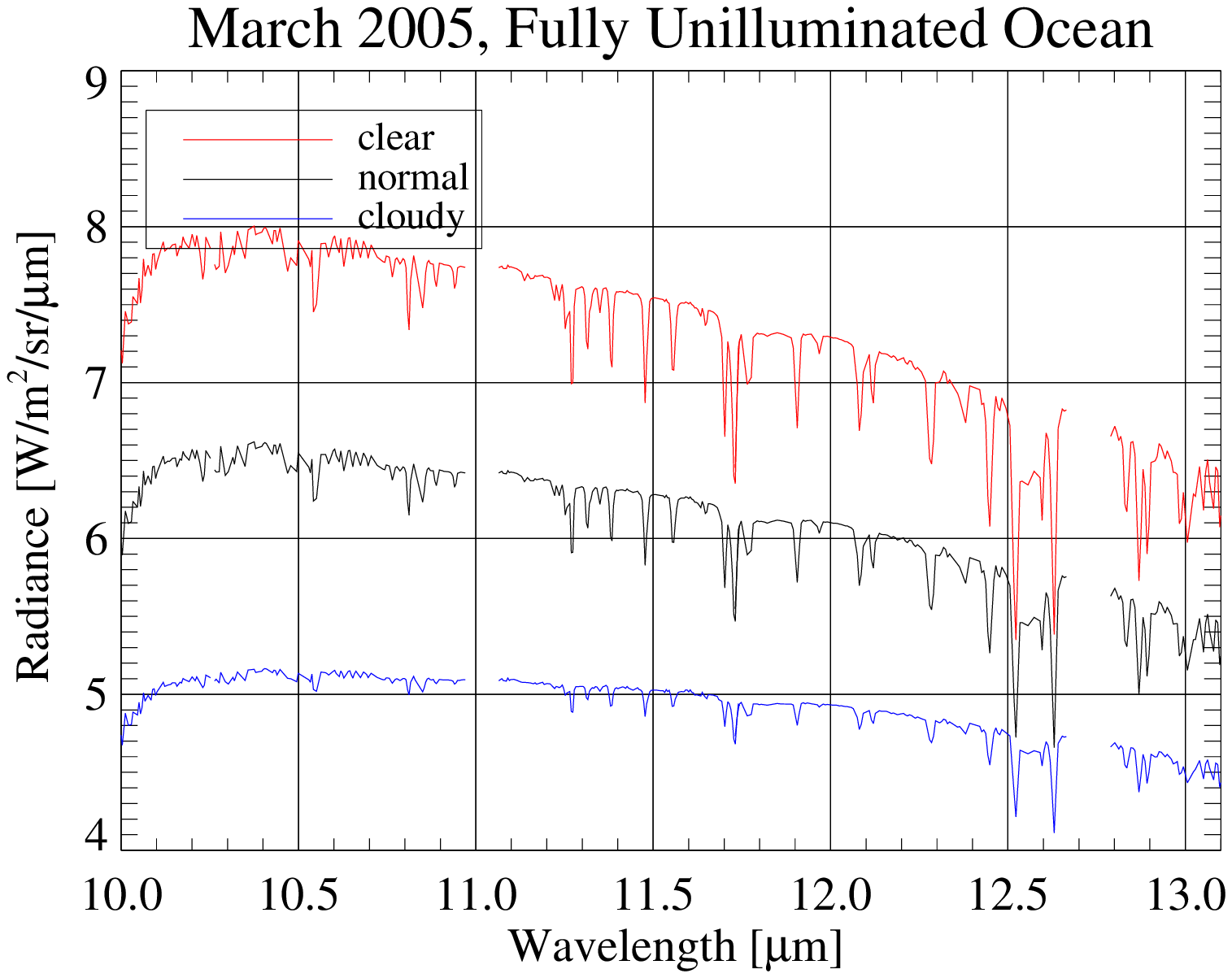}
\includegraphics[width=.32\textwidth]{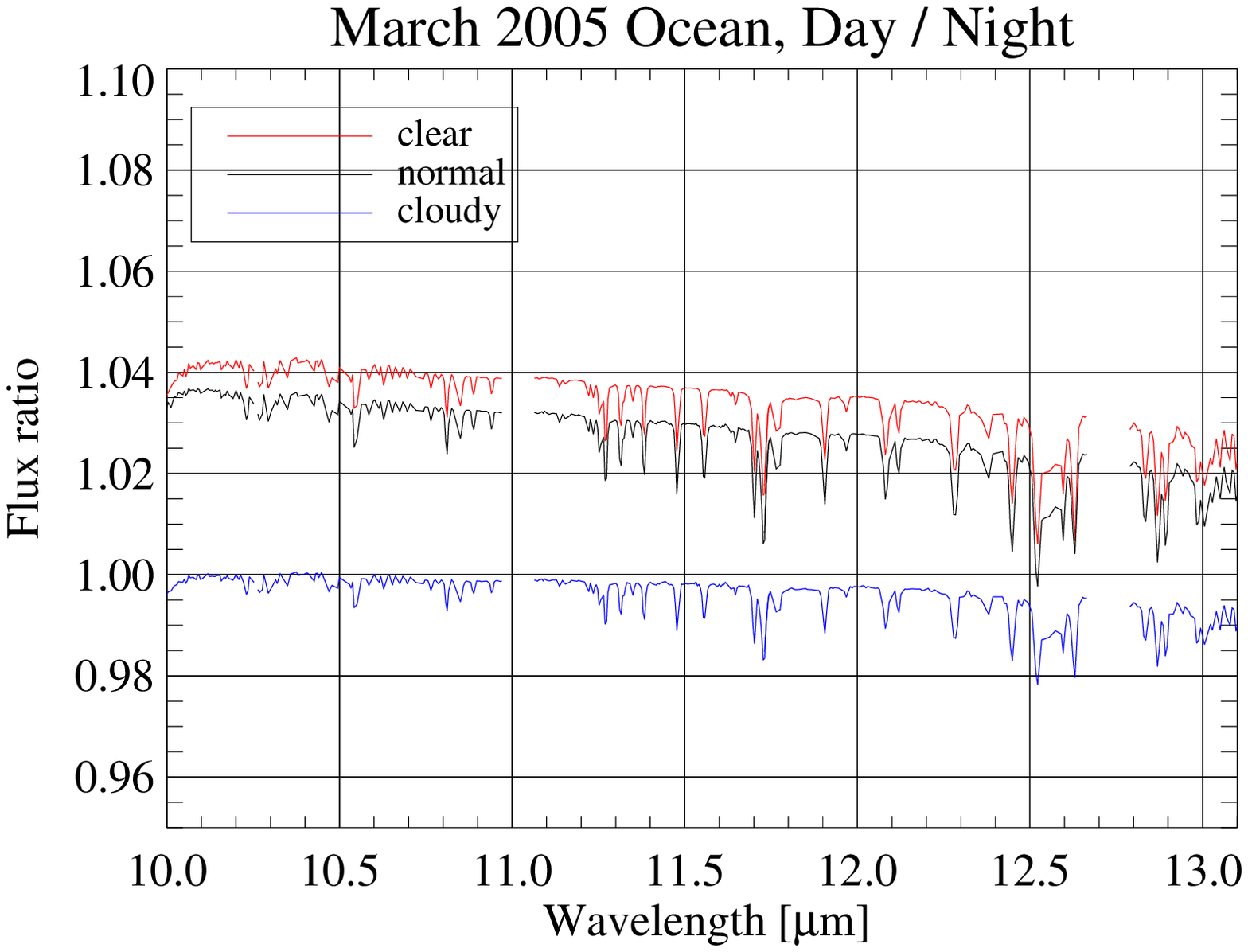} \\
\includegraphics[width=.32\textwidth]{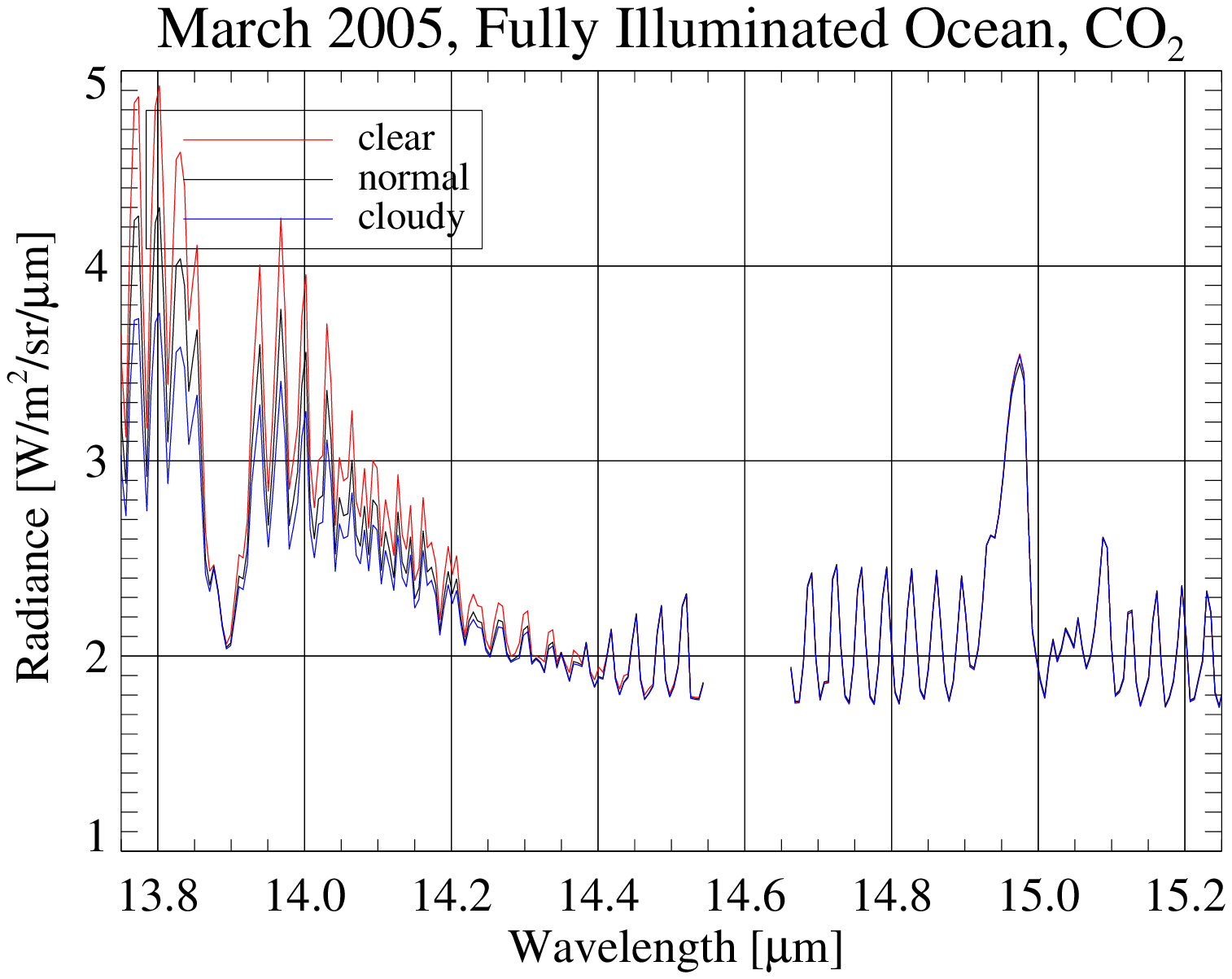}
\includegraphics[width=.32\textwidth]{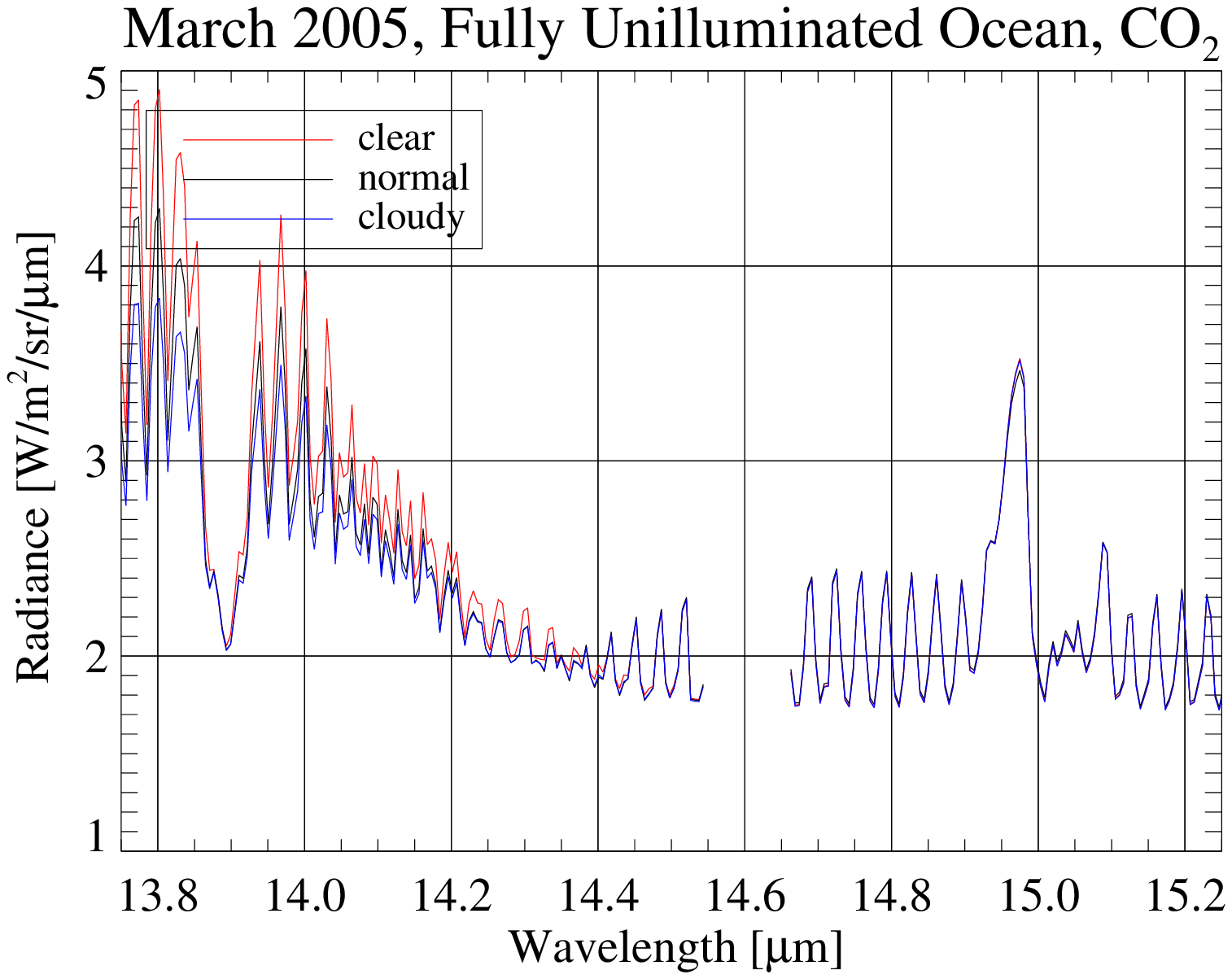}
\includegraphics[width=.32\textwidth]{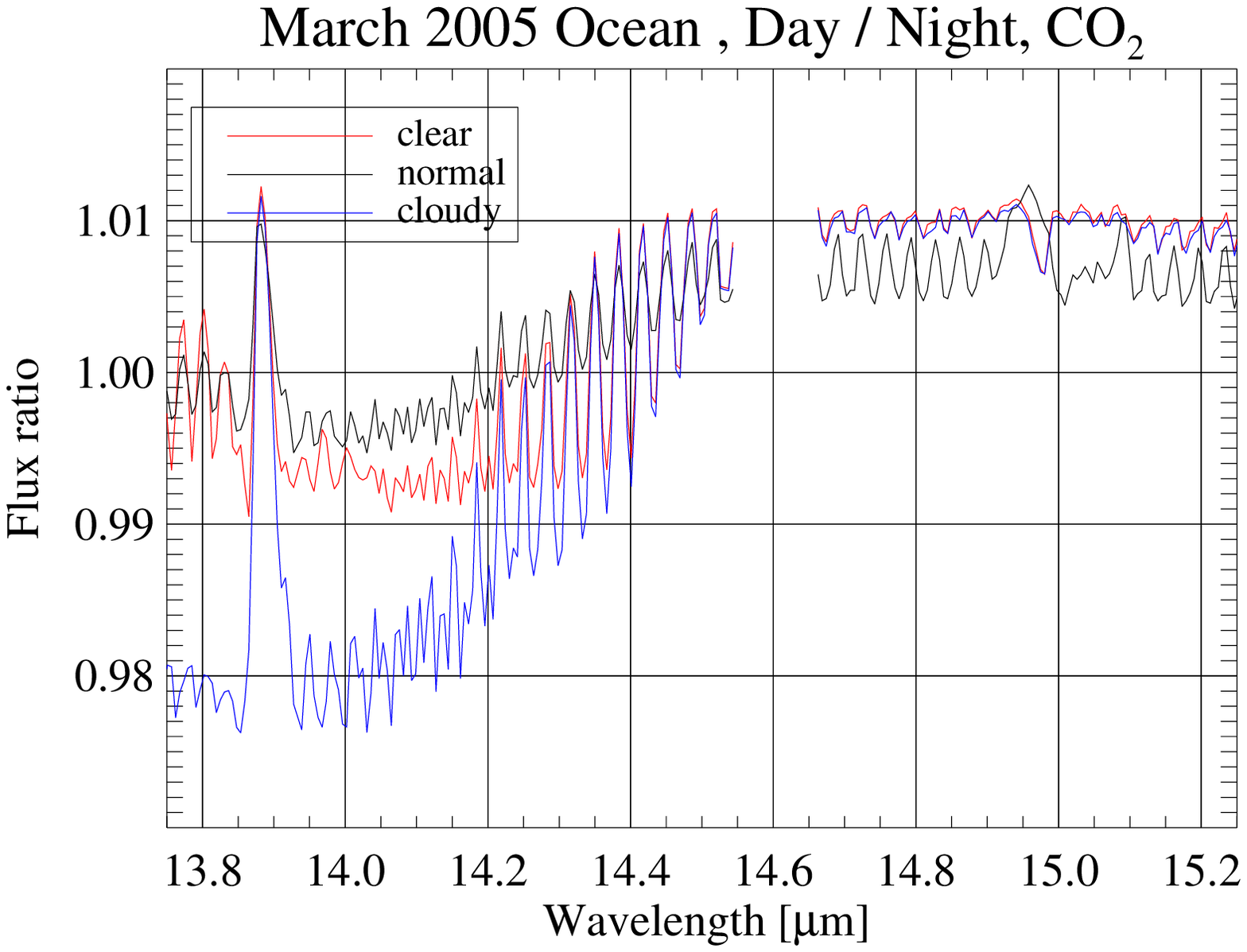}
\caption{Five spectral regions (from top to bottom) are shown for disk averaged spectra
of an edge on view of Earth that is (1) fully illuminated ``day'' and (2) unilluminated ``night''
ocean views and (3) the ratio between day and night (from left to right).
The weak lines between 3.8 and 4.0 $\mu m$ are due to HDO and the large ratio between
day and night in this region is due to reflected solar radiation which is more significant for
the cloudy cases.  Absorption lines of CO$_2$, O$_3$, H$_2$O, and CH$_4$ are
clearly seen. The CO$_2$ and O$_3$ emission lines are due to stratospheric
emission.
The lines from 10-13 $\mu m$ are due to H$_2$O and the slope in
this region may be an indicator of water even if the lines are not resolved.}
\label{ocean_spec_zoom}
\end{figure}

Figure~\ref{ocean_spec_zoom} shows the fully illuminated and unilluminated Earth spectra
of the H$_2$O,  O$_3$, CO$_2$, and CH$_4$ spectral features and the ratio of the spectral feature
in the daytime and the night time.
The 14~$\mu$m CO$_2$ feature is less sensitive to day night difference and varying
cloud amounts than the 4 $\mu$m CO$_2$ lines that include some reflected sunlight.
However, the saturated line cores of O$_3$ and CO$_2$ due to the warm stratosphere are
similar for both the clear and cloudy cases.  Therefore, if an earth-like extrasolar
planet were completely covered by clouds we may still be able to detect O$_3$ emission
if the telescope has sufficient spectral resolution.

Since the first generation of telescopes capable of directly detecting spectra of extrasolar
terrestrial planets will likely not have the spectral resolution of AIRS many of the
features seen in the spectra described in this paper may not be observable.
The space-based NASA-Terrestrial Planet Finder mission concepts (TPF-O,
TPF-C, TPF-I) or the ground-based European Extremely Large Telescope
(E-ELT) are expected to image and characterize exoplanets down to
Super-Earth and Earth-size. In these cases, the spectral resolution
obtainable will be limited by the photon noise, so a reasonable estimate is
a spectral resolution of $\sim$ 50 for super-Earths at 10 pc.  Higher spectral
resolution spectra in the Infrared, though, could be obtained with JWST for super-Earths
in the habitable zone of M stars using the secondary transit technique.

\begin{figure}
\includegraphics[width=.9\textwidth]{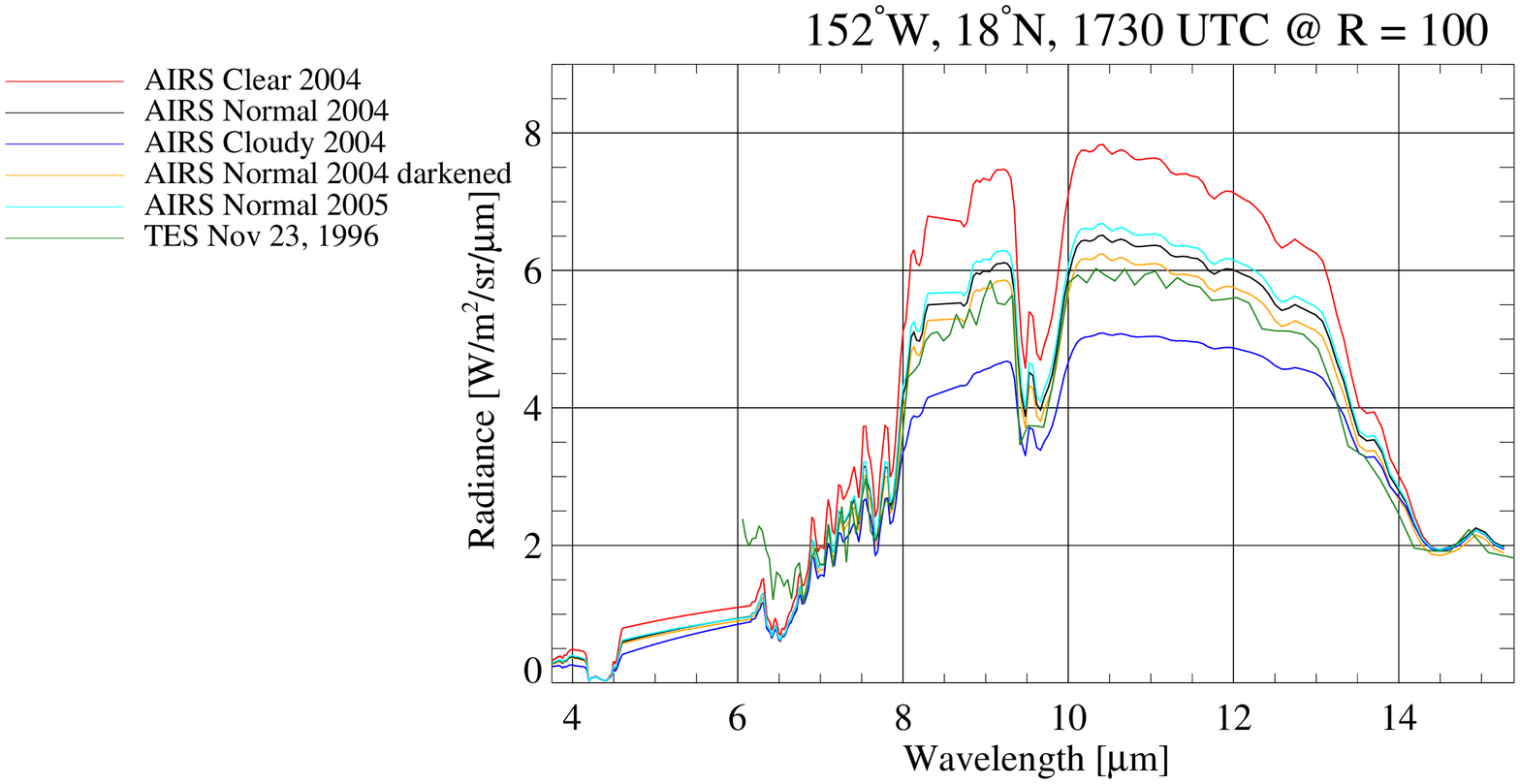}
\includegraphics[width=.9\textwidth]{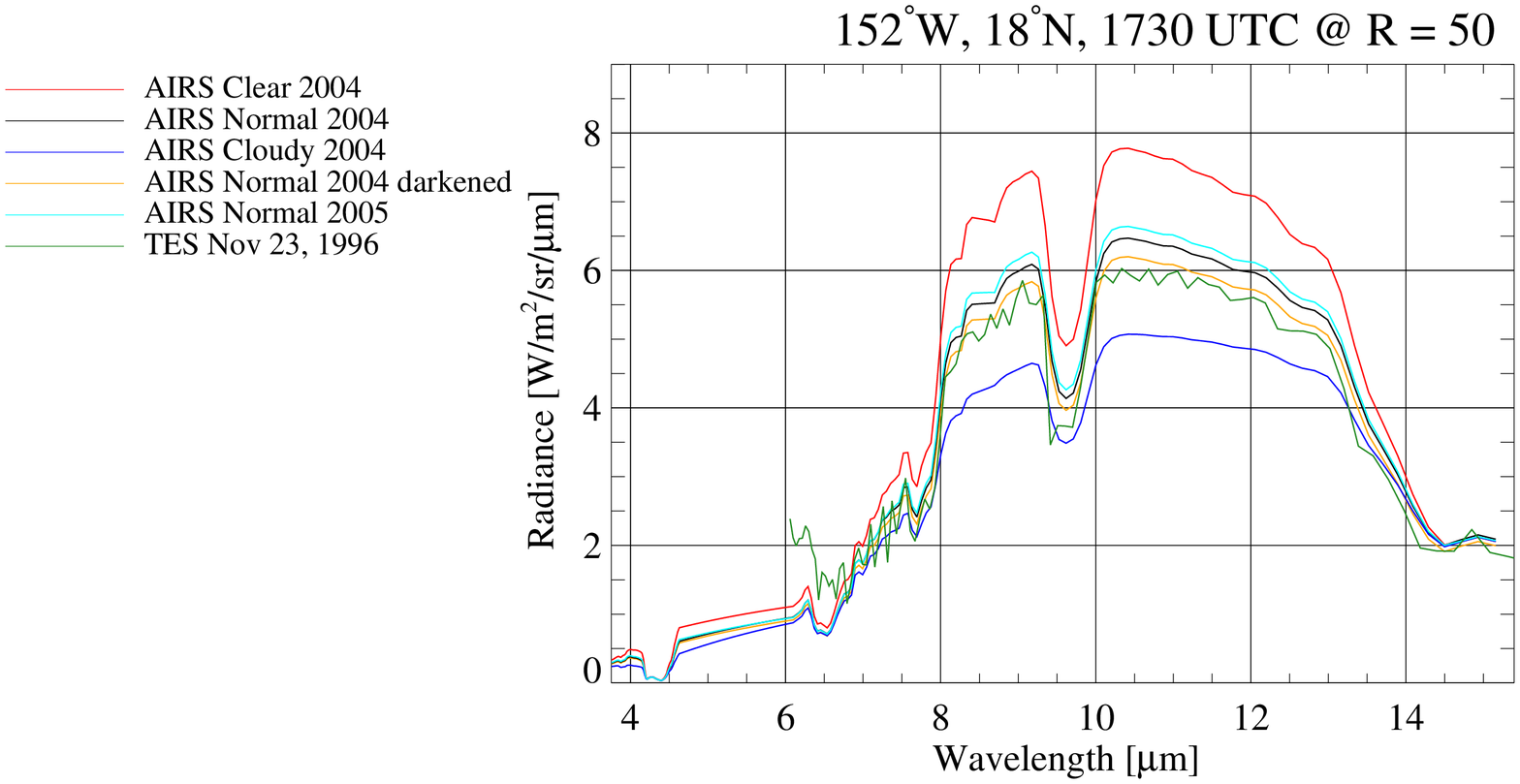}
\includegraphics[width=.9\textwidth]{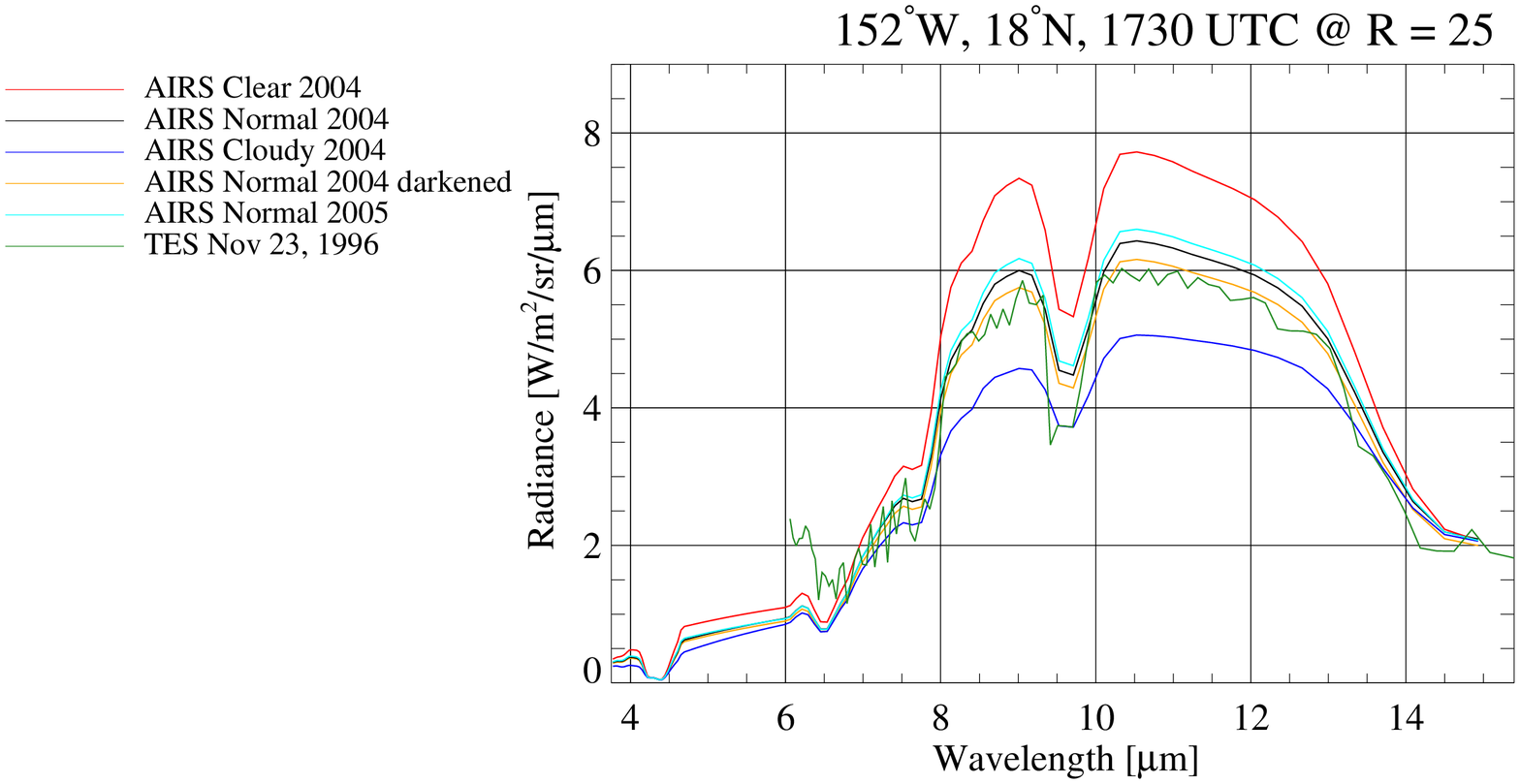}
\caption{The 3 panels show the same as Figure~\ref{tescomparison} with the spectral resolution degraded to R = 100, 50, and 25 (from top to bottom).}
\label{degradedspec}
\end{figure}

Figure~\ref{degradedspec} displays AIRS disk averaged
spectra degraded to spectral resolutions of R = 100, 50, and 25.  Molecular
absorption lines of H$_2$O,  O$_3$, CO$_2$, and CH$_4$ are
detectable in all but the most cloudy cases.  However, the emission lines due to
stratospheric O$_3$ and CO$_2$ are only detectable in the spectra
with R $=$ 100.

\clearpage

\clearpage

\section{Rotational Variations}
\label{rotational}

Figure~\ref{march_spec} has already shown that fully illuminated ``daytime''
spectra show larger rotational variations depending upon the amount of land
in the field of view
and the daytime absorption features are deeper than in the
unilluminated ``nighttime'' spectra.  However, telescopes like
TPF and Darwin will not
observe planets that are fully illuminated or fully unilluminated
since the planets must be away from their star.
A more likely observing scenario would be for the planet to be partially
illuminated.  Thus, we examine the rotational radiance variations due to planet rotation
by creating disk averaged spectra for edge-on and pole-on viewing geometries
in two hour increments (UT = 0--23 hours).

\begin{figure}
\begin{center}
\includegraphics[width=.99\textwidth]{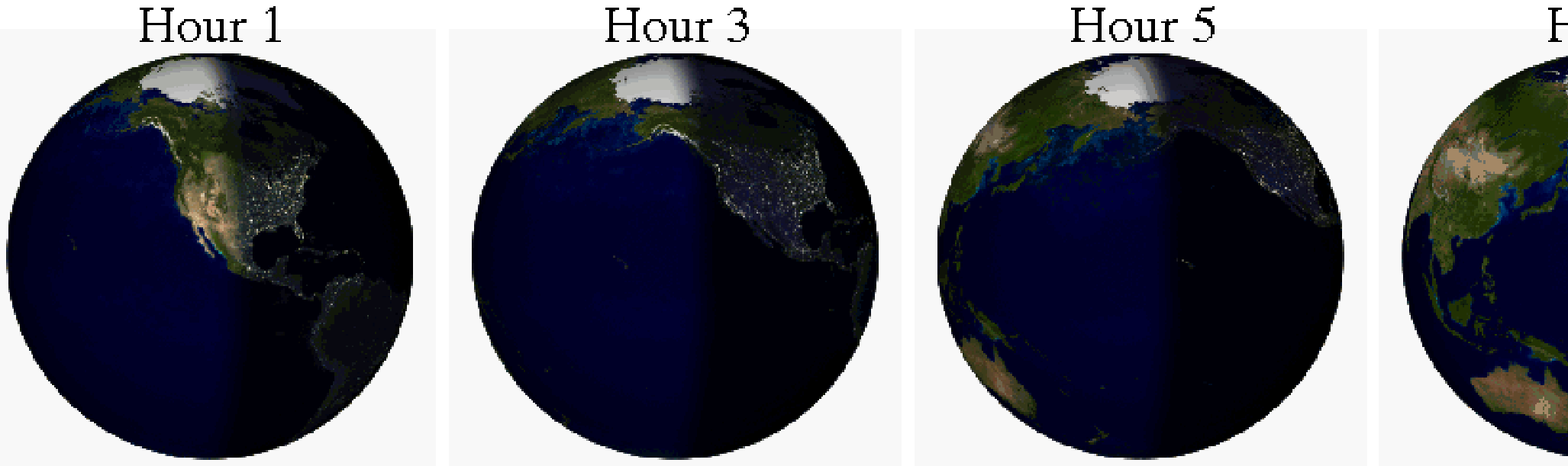} \\
\includegraphics[width=.6\textwidth]{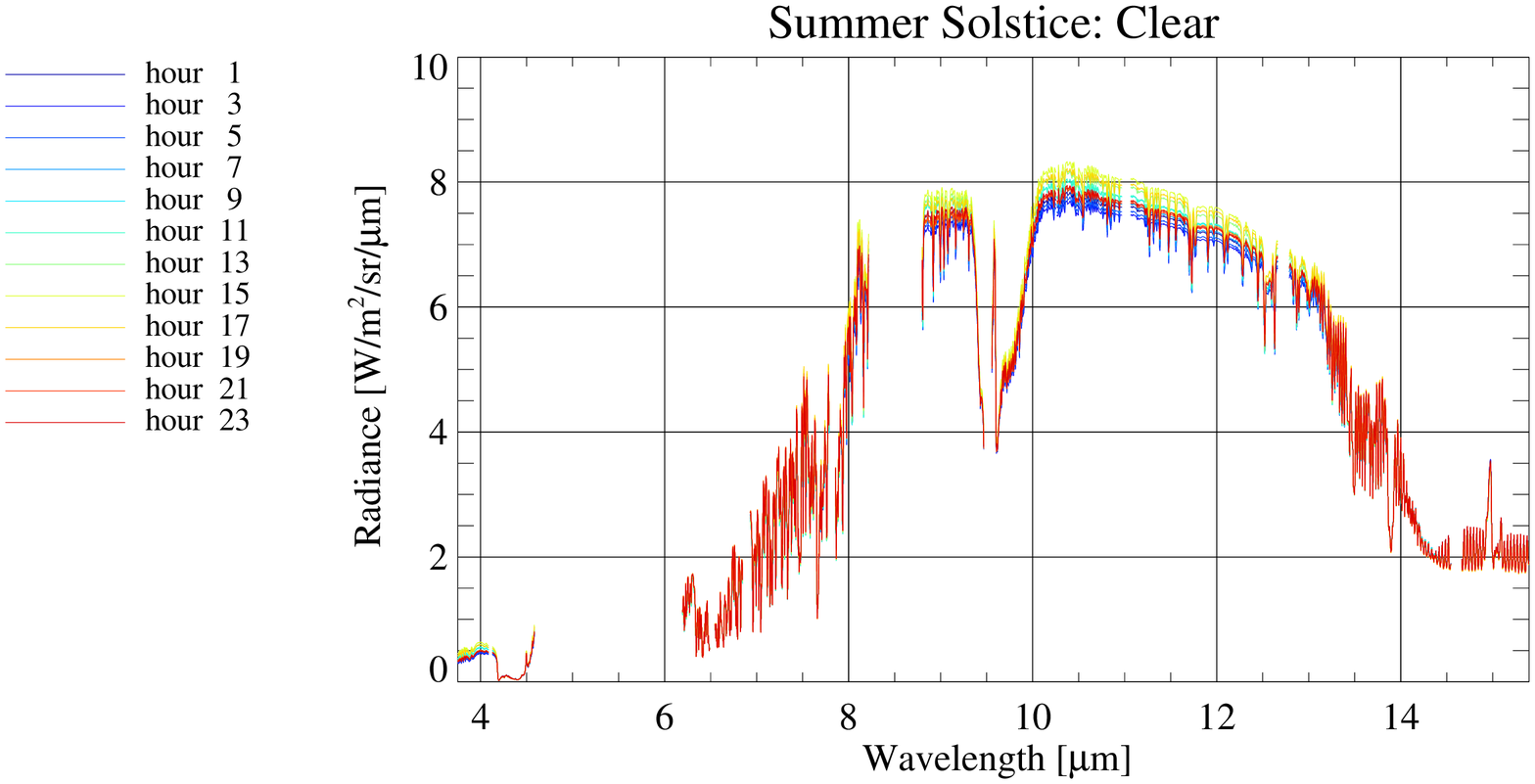} \\
\includegraphics[width=.6\textwidth]{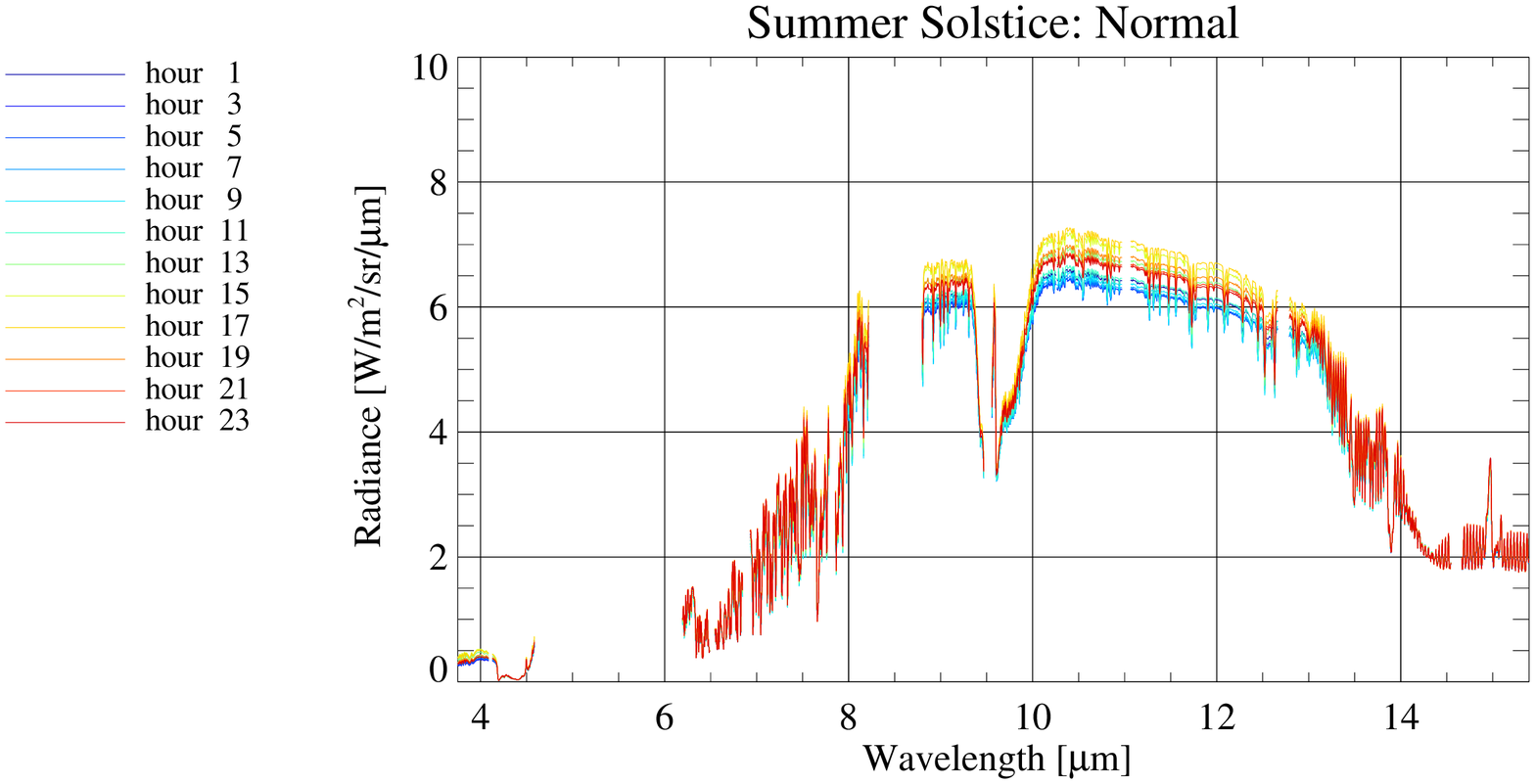} \\
\includegraphics[width=.6\textwidth]{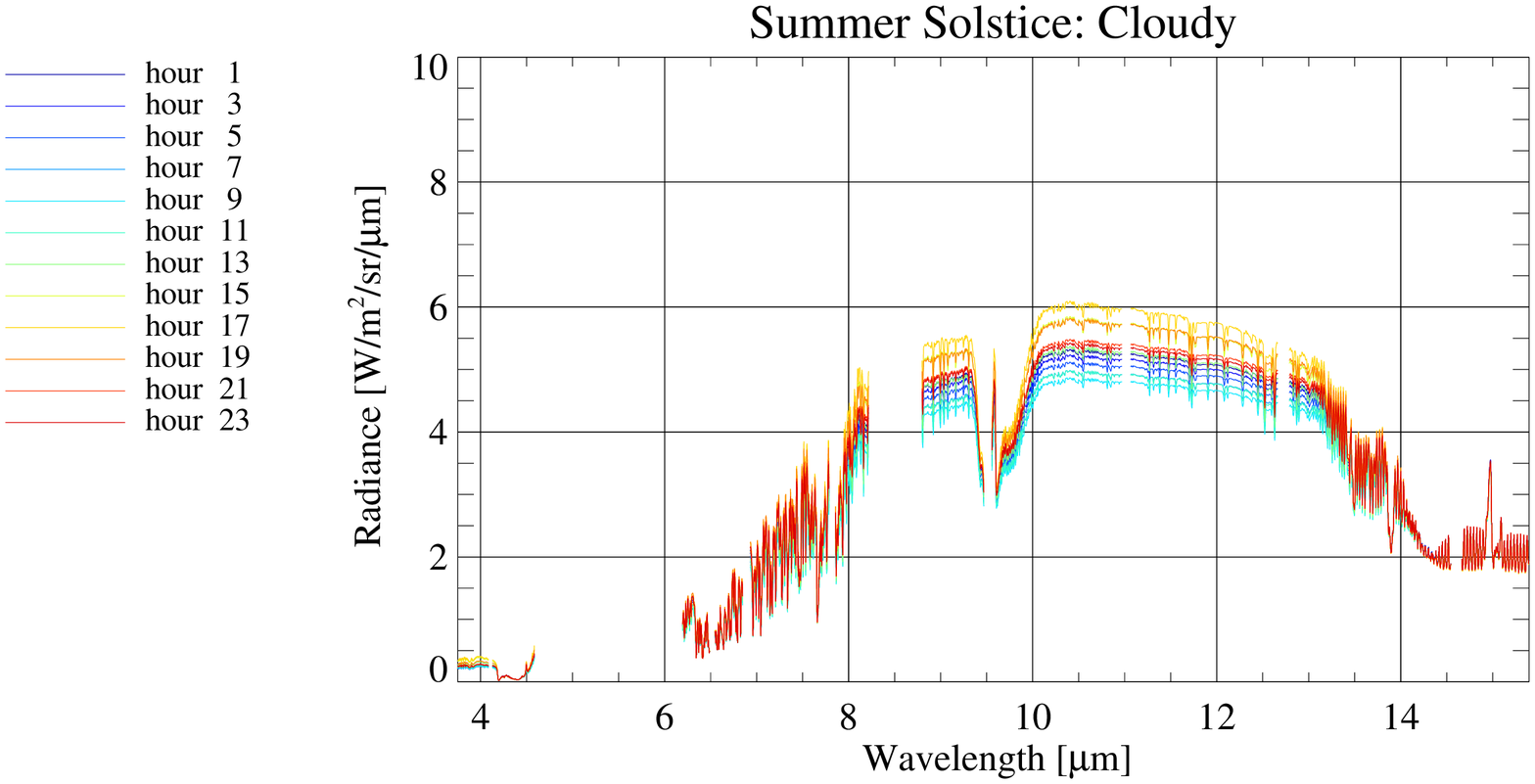}
\end{center}
\caption{Infrared spectra are displayed for 
clear, normal, and cloudy scenes for an edge-on viewing geometry on March 26, 2005.}
\label{fig:ir_edgeviews_mar}
\end{figure}

Figure~\ref{fig:ir_edgeviews_mar} shows an edge-on view of Earth seen
in March from a distant observer located at  RA = 6.0 hours \& Dec = 23.5$^{\circ}$,
the ``Summer Solstice'' view. The spectral features (e.g., O$_3$, H$_2$O, CO$_2$, and CH$_4$) are visible at
all times and the amplitude of the variability
in the continuum is $\sim$ 10\% as calculated by \citet{des02}.  The magnitude of the
radiometric variability is probably exaggerated for the cloudy cases since some clear scenes
over the Sahara were misidentified as cloudy.

\begin{figure}
\begin{center}
\includegraphics[width=.99\textwidth]{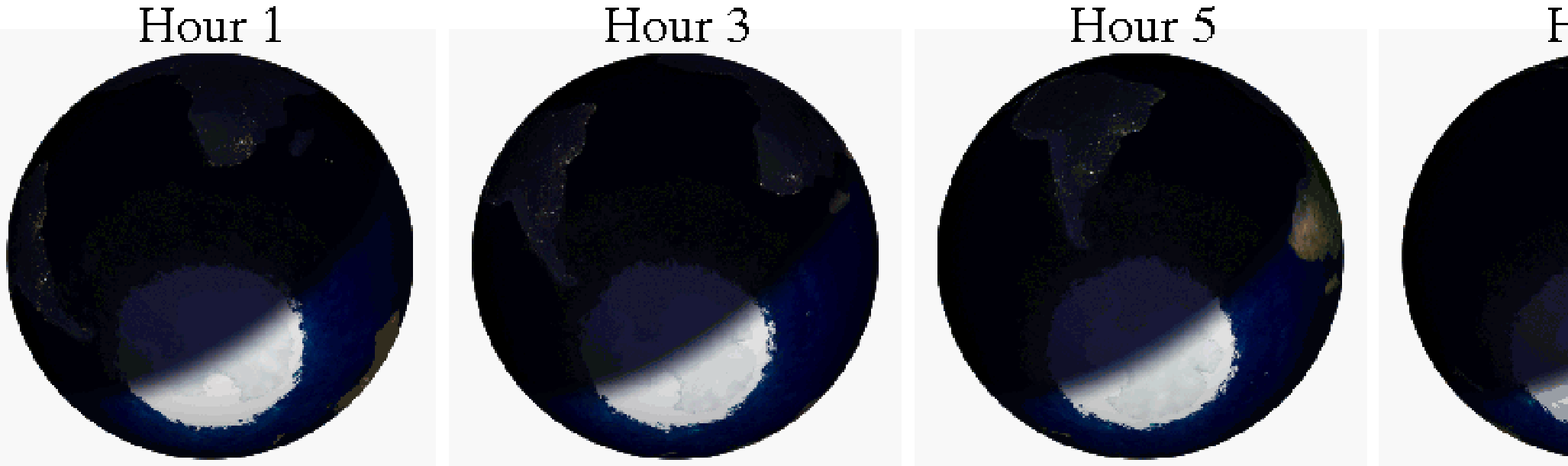} \\
\includegraphics[width=.6\textwidth]{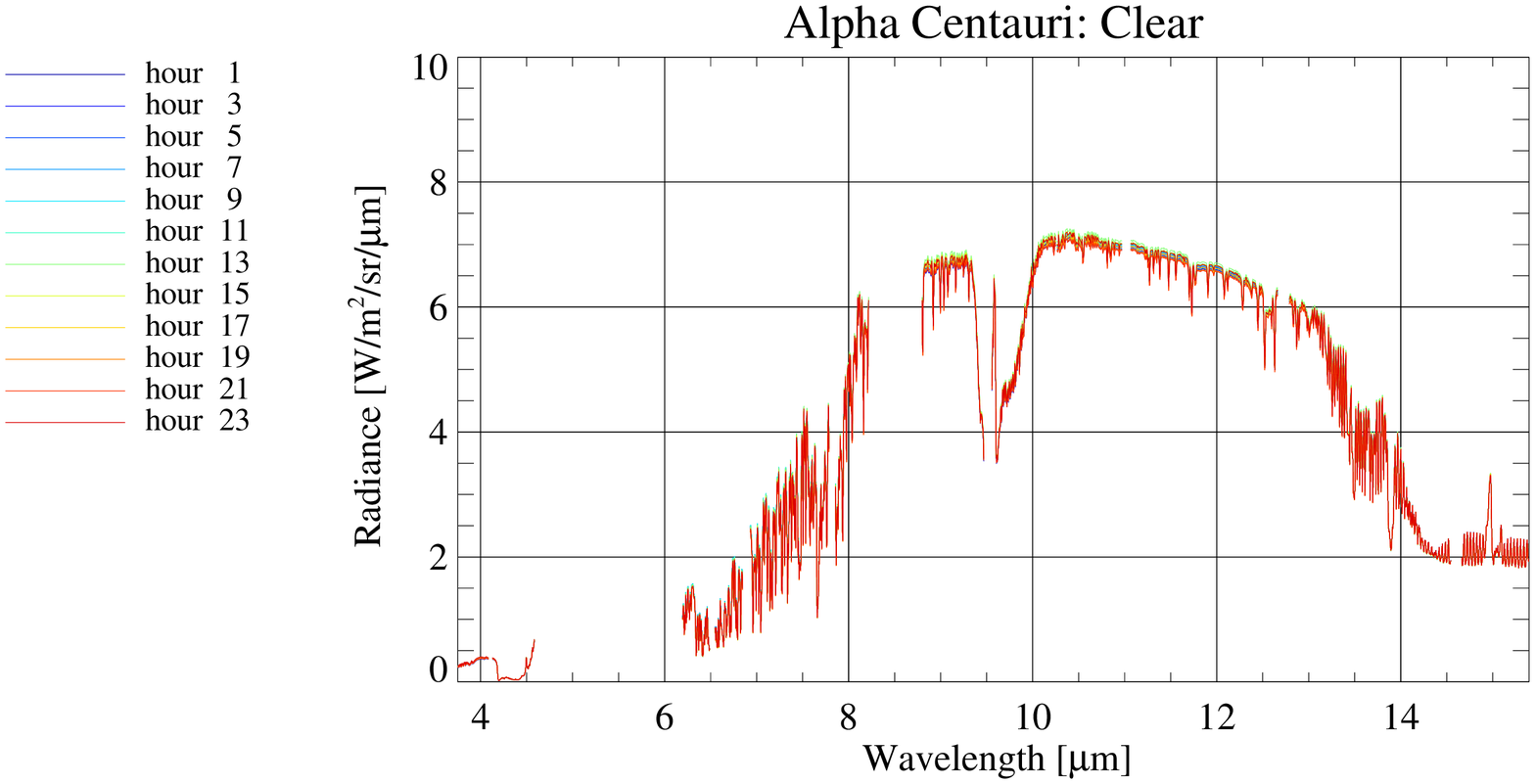} \\
\includegraphics[width=.6\textwidth]{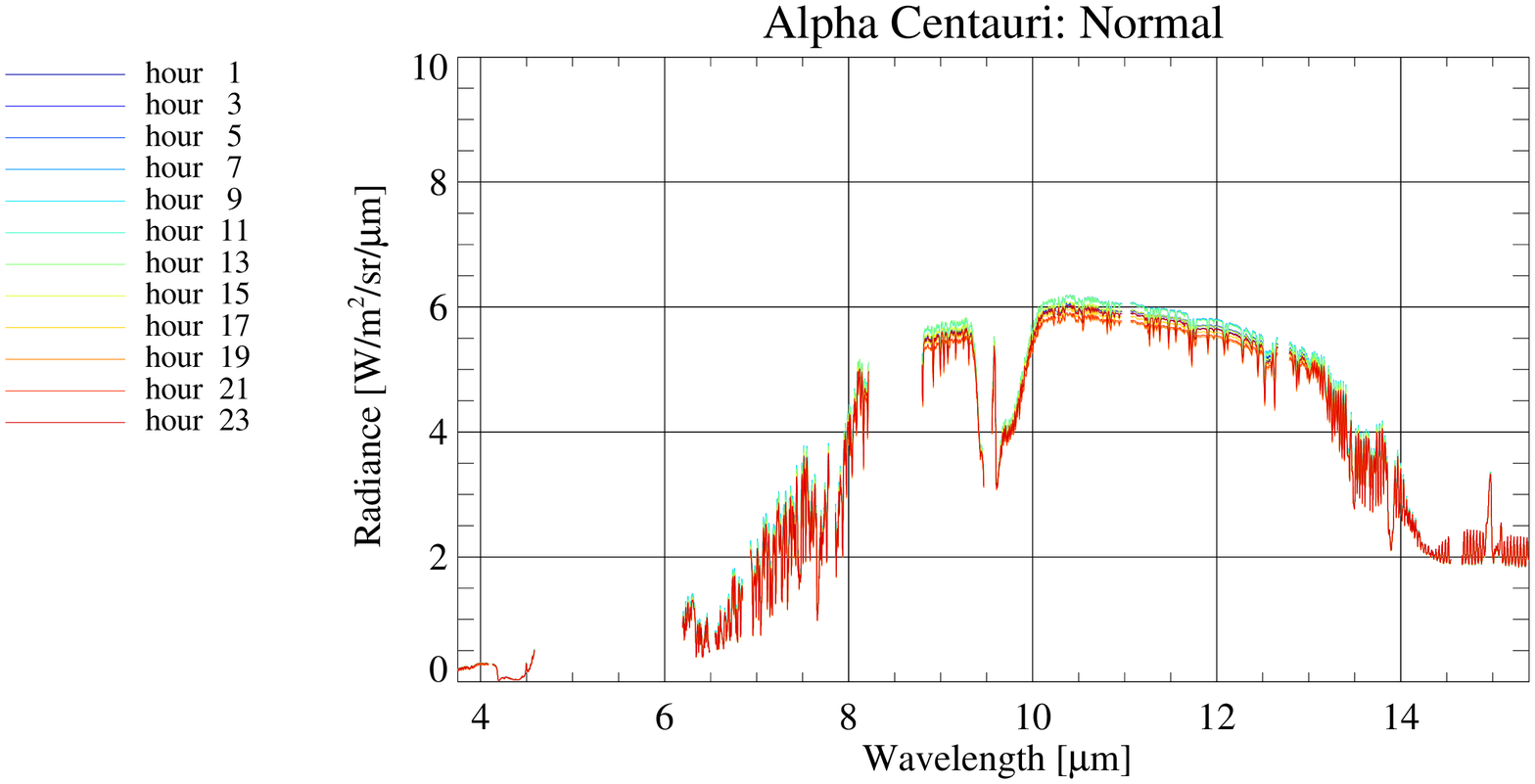} \\
\includegraphics[width=.6\textwidth]{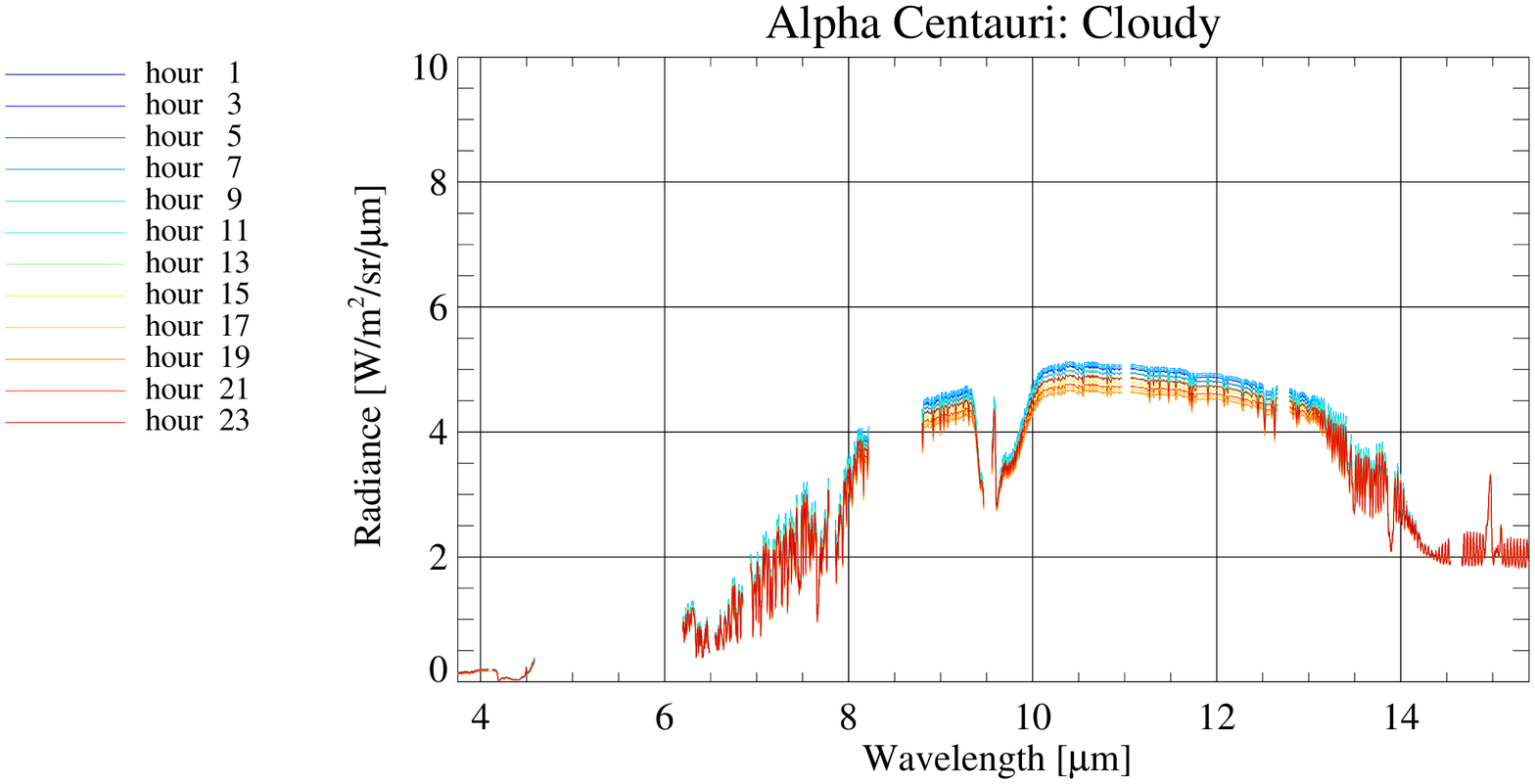} \\
\end{center}
\caption{Infrared spectra are displayed for 
clear, normal, and cloudy scenes for a nearly pole-on viewing geometry of Earth viewed
from $\alpha$ Centauri on March 26, 2005.}
\label{fig:ir_polarviews_mar}
\end{figure}

Figure~\ref{fig:ir_polarviews_mar} shows a nearly pole-on view of Earth as seen
by an observer located at RA = 14.6 hours \& Dec = $-$60.63$^{\circ}$, the ``$\alpha$ Centauri'' view.
The same spectral lines are visible but the spectra are colder and the amplitude
of the radiometric variability is reduced.  A similar pole-on view of the northern
hemisphere shows a larger rotational variability because of the larger fraction of land
in the northern hemisphere.

\clearpage
\pagestyle{empty}
\setlength{\voffset}{-12mm}

\clearpage
\pagestyle{plaintop}
\setlength{\voffset}{0mm}

\section{Seasonal Variations}
\label{seasonal}

Although the whole seasonal cycle will not be observable by TPF or Darwin for edge-on views
because the planet will be too close to the star for some cases (e.g., eclipsing planets)
spectra may be obtained by other means.  Specifically the transit method makes it
possible to study the atmosphere of the transiting planet. In particular, the secondary
eclipse (when the planet is blocked by its star) allows a direct measurement of the planet's
radiation. If the star's contribution during the secondary eclipse is
subtracted from the observed spectrum before or after, only the signal
caused by the planet remains. It is then possible to measure the planet's
emitted \citep{deminga,charbonneaua} or reflected
spectrum \citep{rowe06} depending on the wavelength region selected
for the observations.  Pole-on views will be observable for all seasons
provided there is sufficient separation between the star and planet.
If the planet is tilted with respect to its orbit and there is an uneven distribution
of land and sea (like Earth) we expect to see seasonal variations in the temperature and perhaps
atmospheric gases.  For a life bearing planet, seasonal variations in O$_3$ \& CO$_2$ may
even be indicators of life. 
We simulated 24 hour integrations by averaging the radiances observed over a 24 hour period
and did this for 12 months from September 2004 -- August 2005.

\begin{figure}
\begin{center}
\includegraphics[width=.99\textwidth]{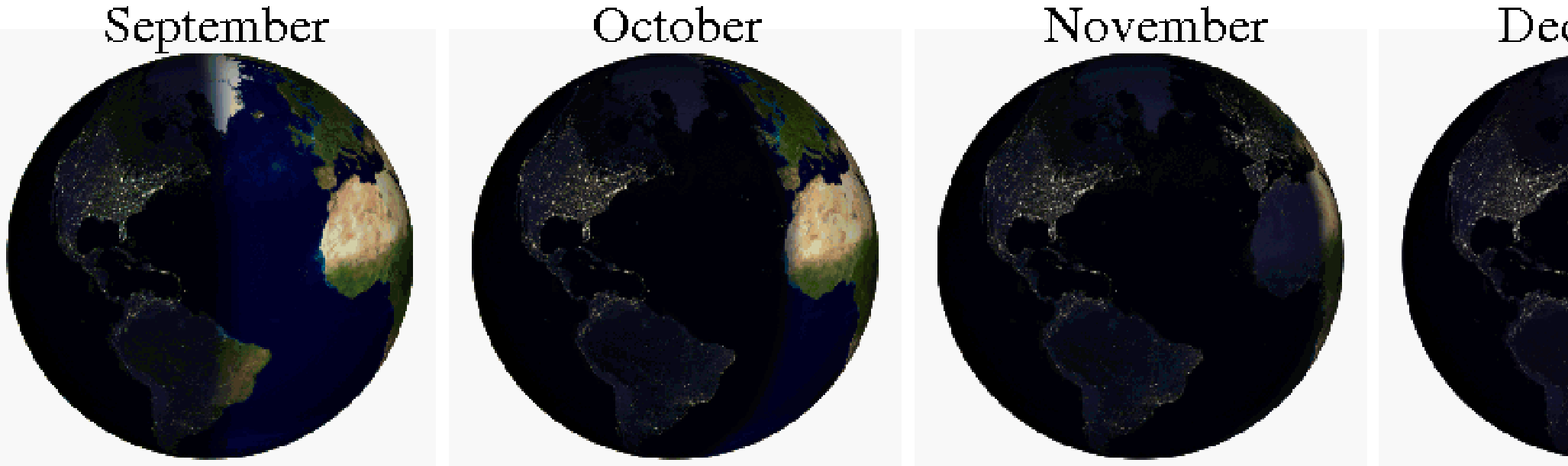} \\
\includegraphics[width=.6\textwidth]{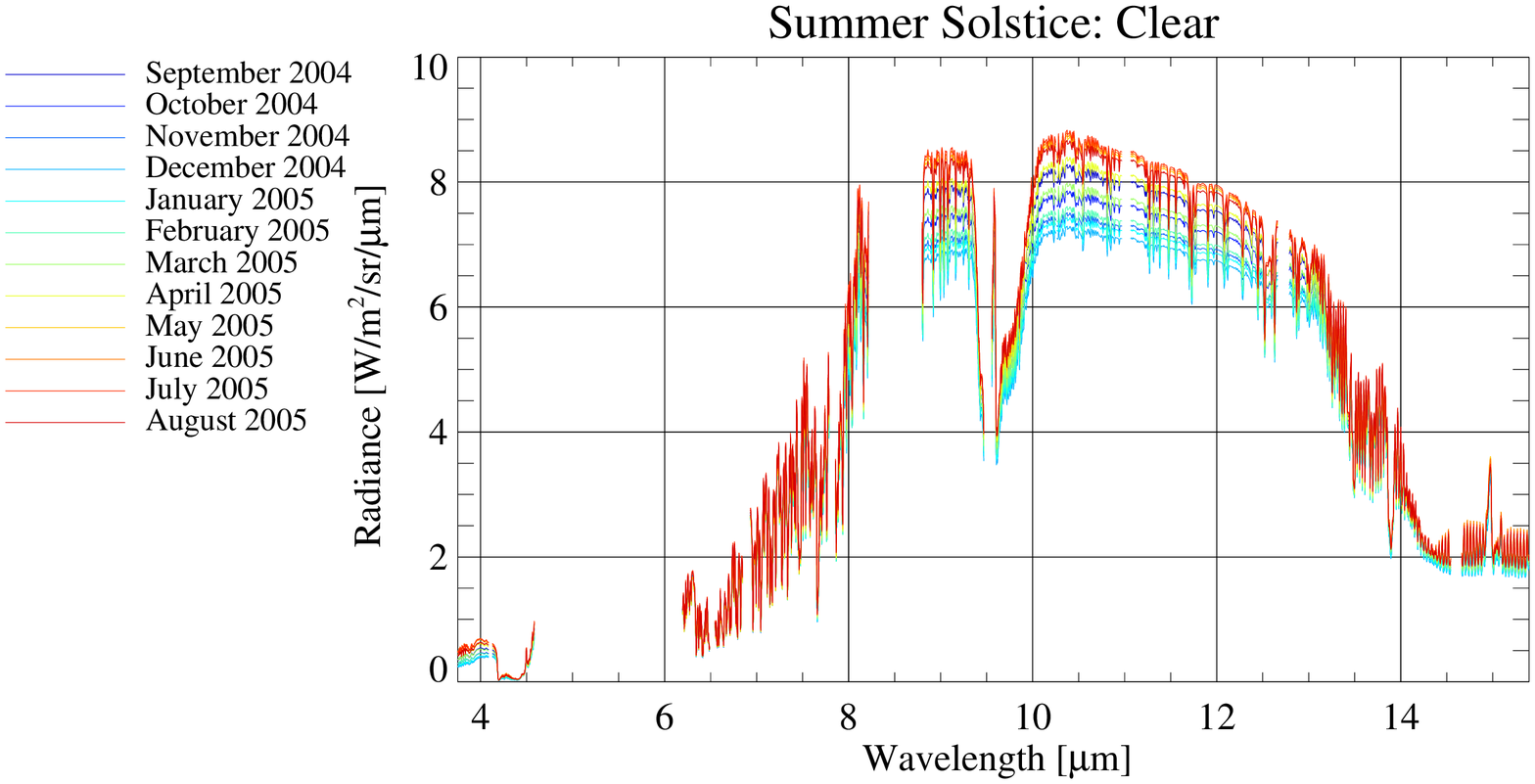} \\
\includegraphics[width=.6\textwidth]{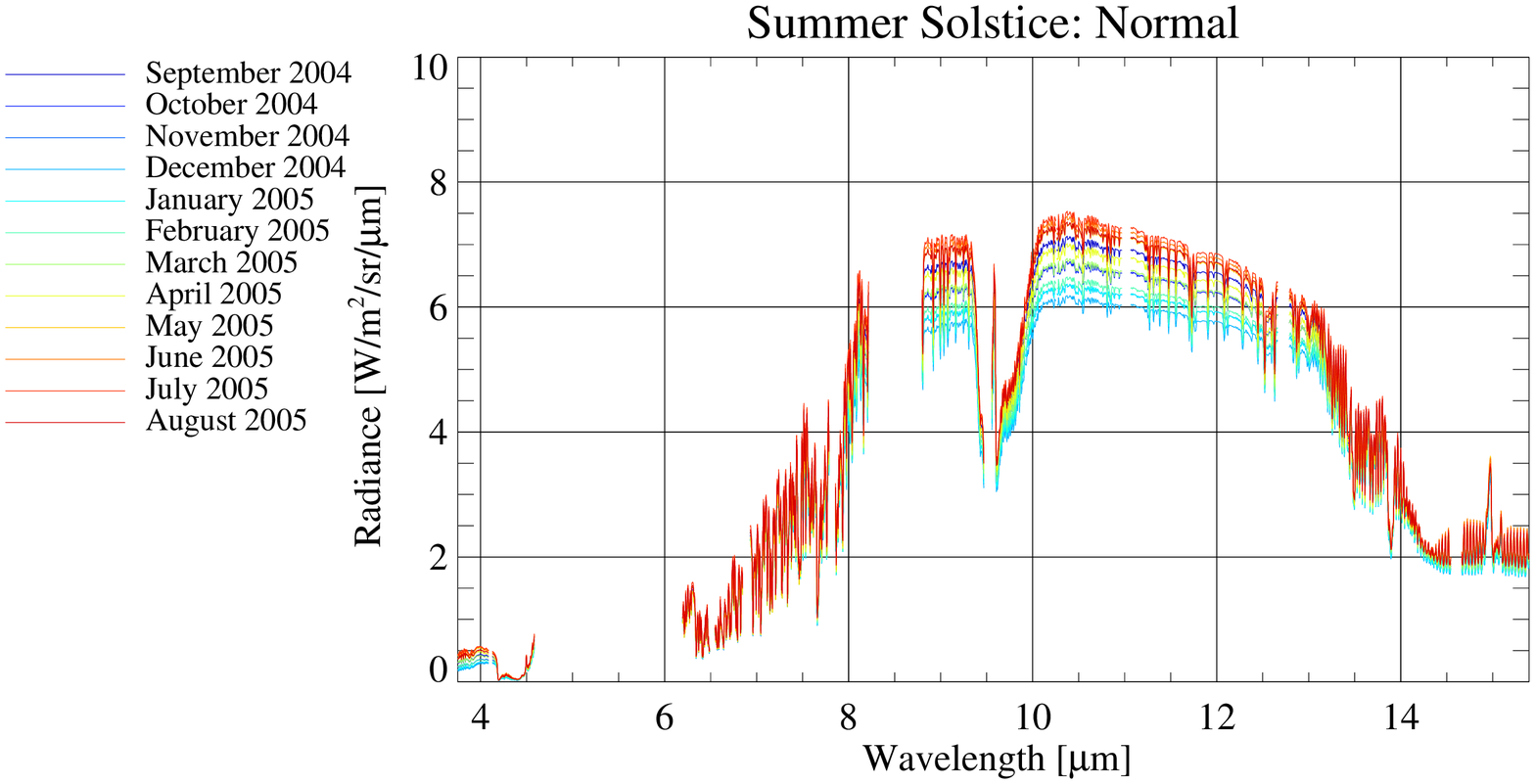} \\
\includegraphics[width=.6\textwidth]{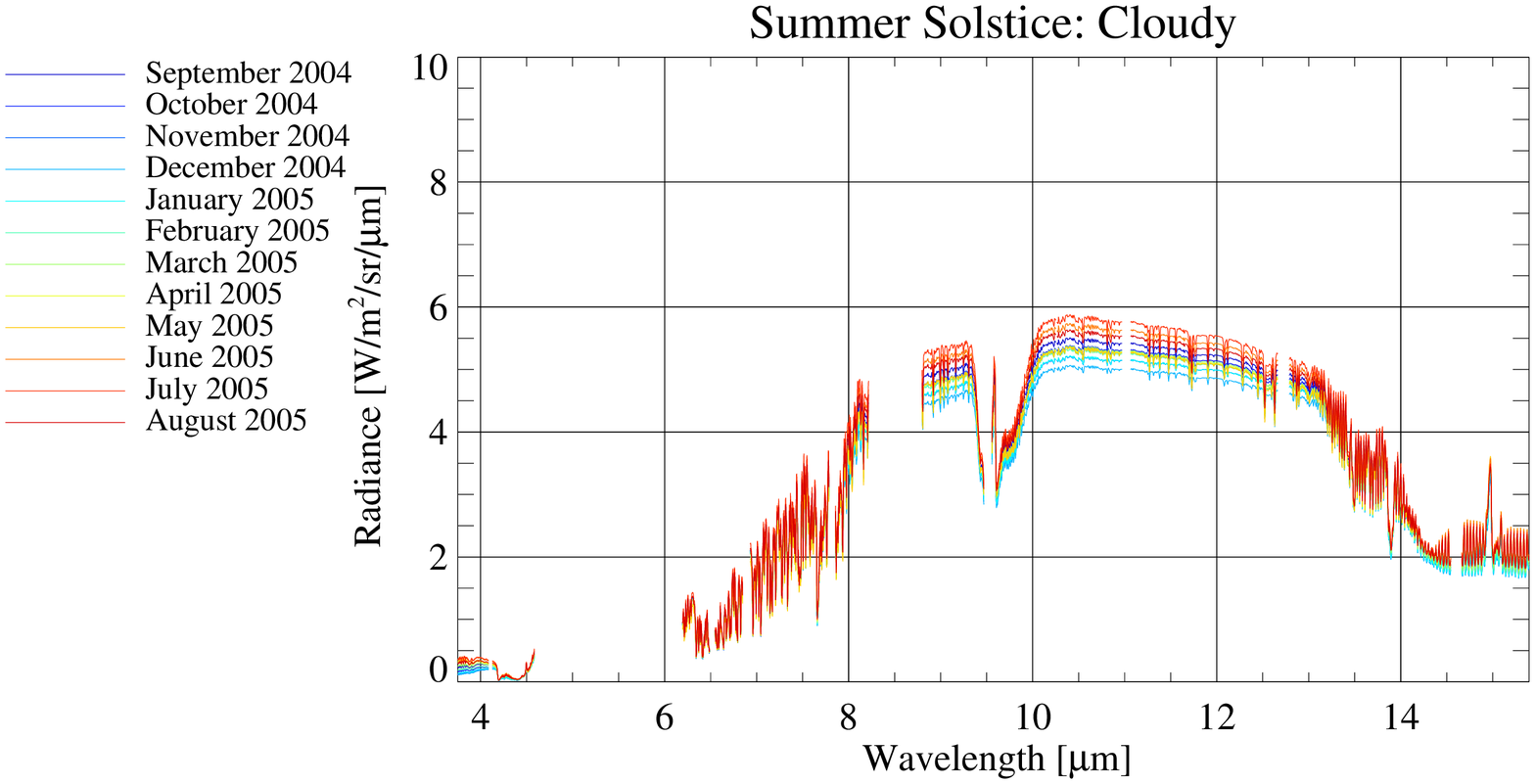}
\end{center}
\caption{The Seasonal variations are displayed for edge-on views of Earth for Clear, Normal, and Cloudy cases.}
\label{fig:ir_edge_seasons}
\end{figure}

Figure ~\ref{fig:ir_edge_seasons} shows the seasonal variations in the mid-infrared for a view of Earth
from RA = 6.0 hours \& Dec = 23.5$^{\circ}$,
the ``Summer Solstice'' view.  The major spectral features in the mid infrared are visible for all seasons
even for the cloud covered cases.
Since this edge-on view is shifted slightly to the northern hemisphere, there is also
a significant seasonal radiometric variability due to the heating of the land in the boreal summer.

\begin{figure}
\begin{center}
\includegraphics[width=.99\textwidth]{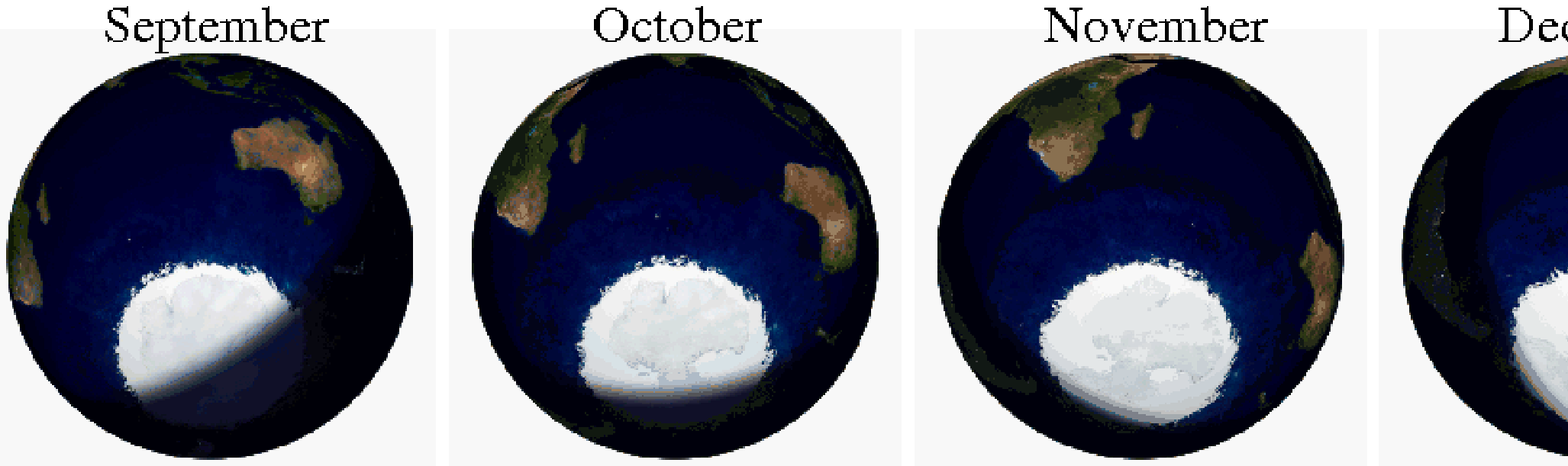} \\
\includegraphics[width=.6\textwidth]{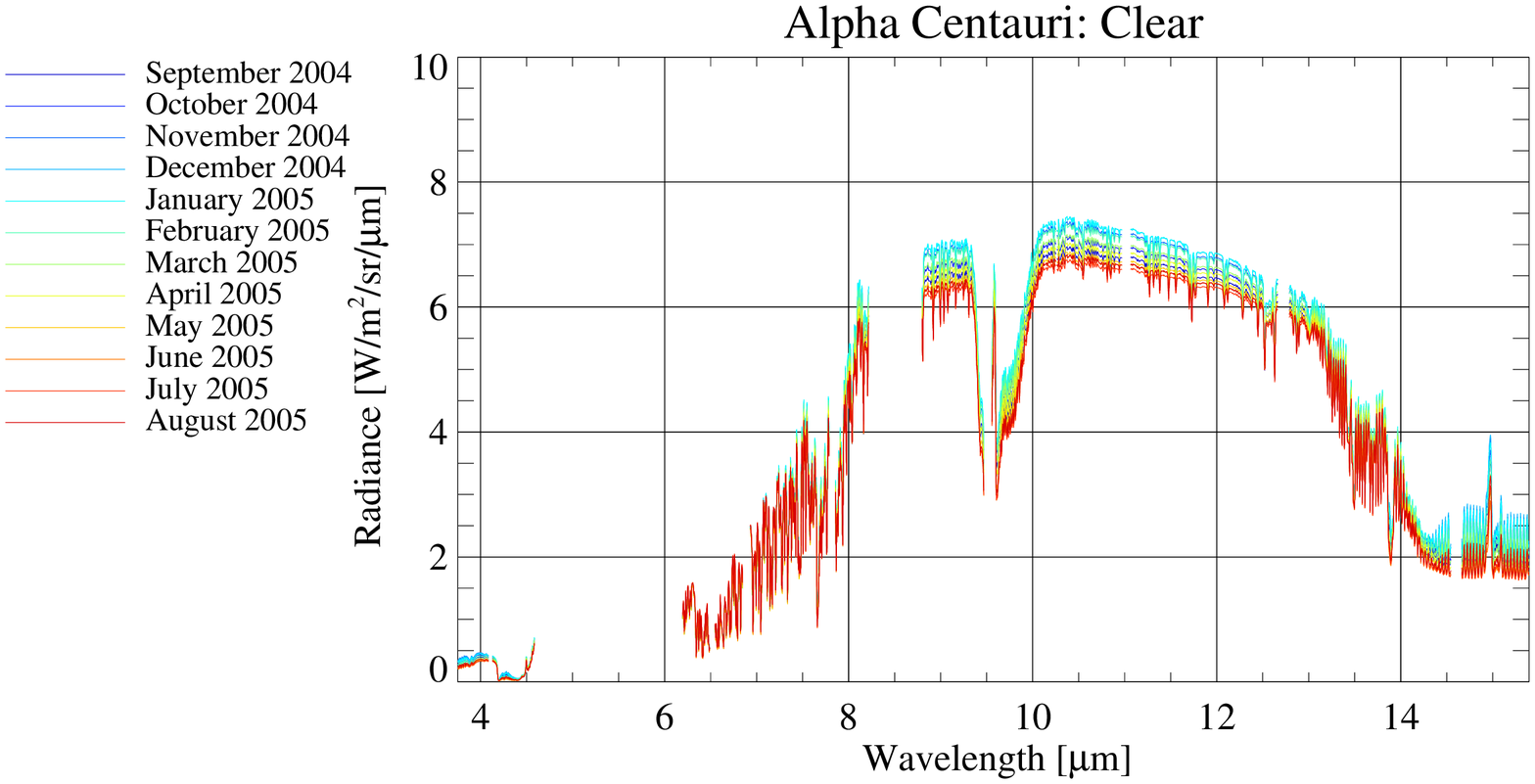} \\
\includegraphics[width=.6\textwidth]{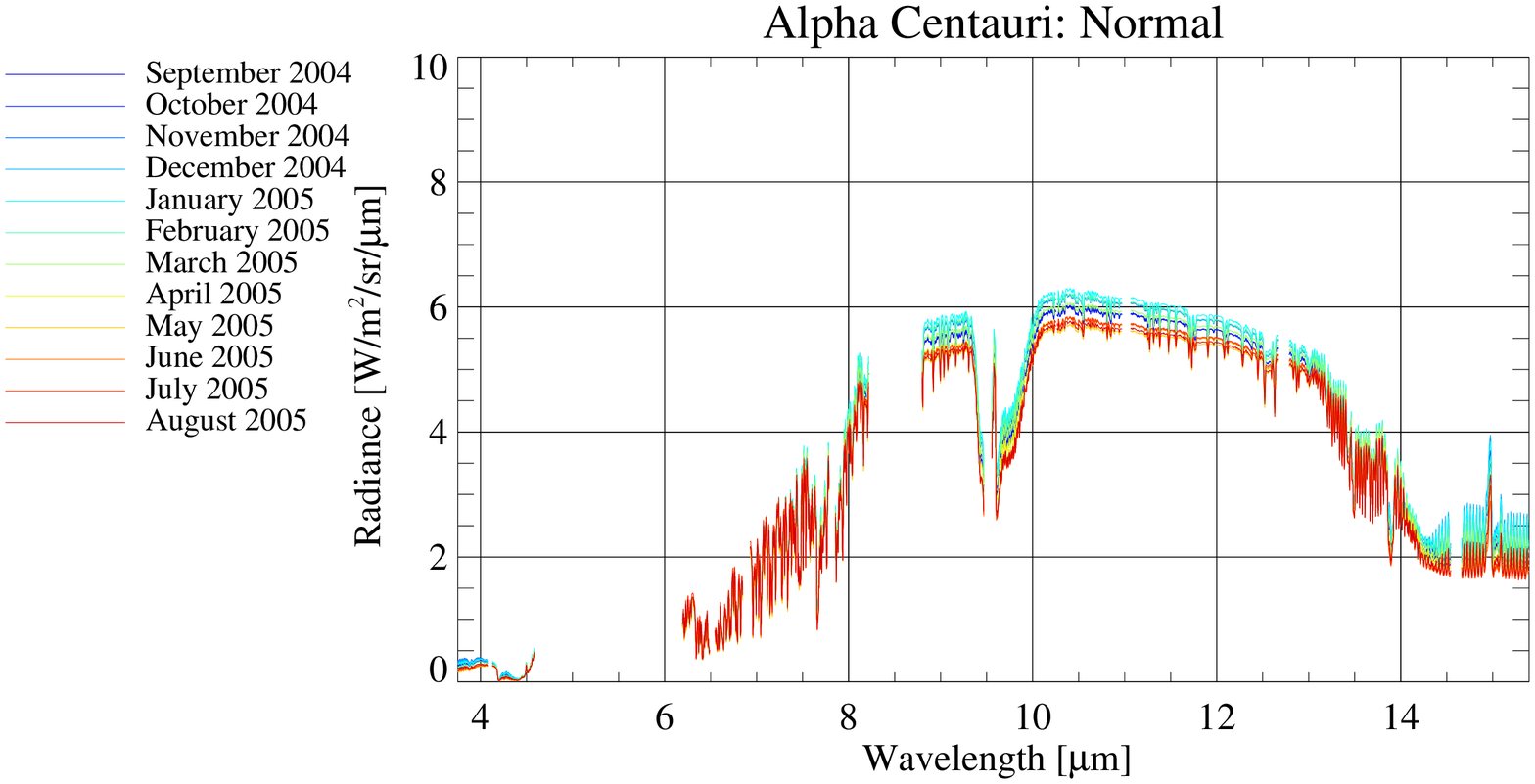} \\
\includegraphics[width=.6\textwidth]{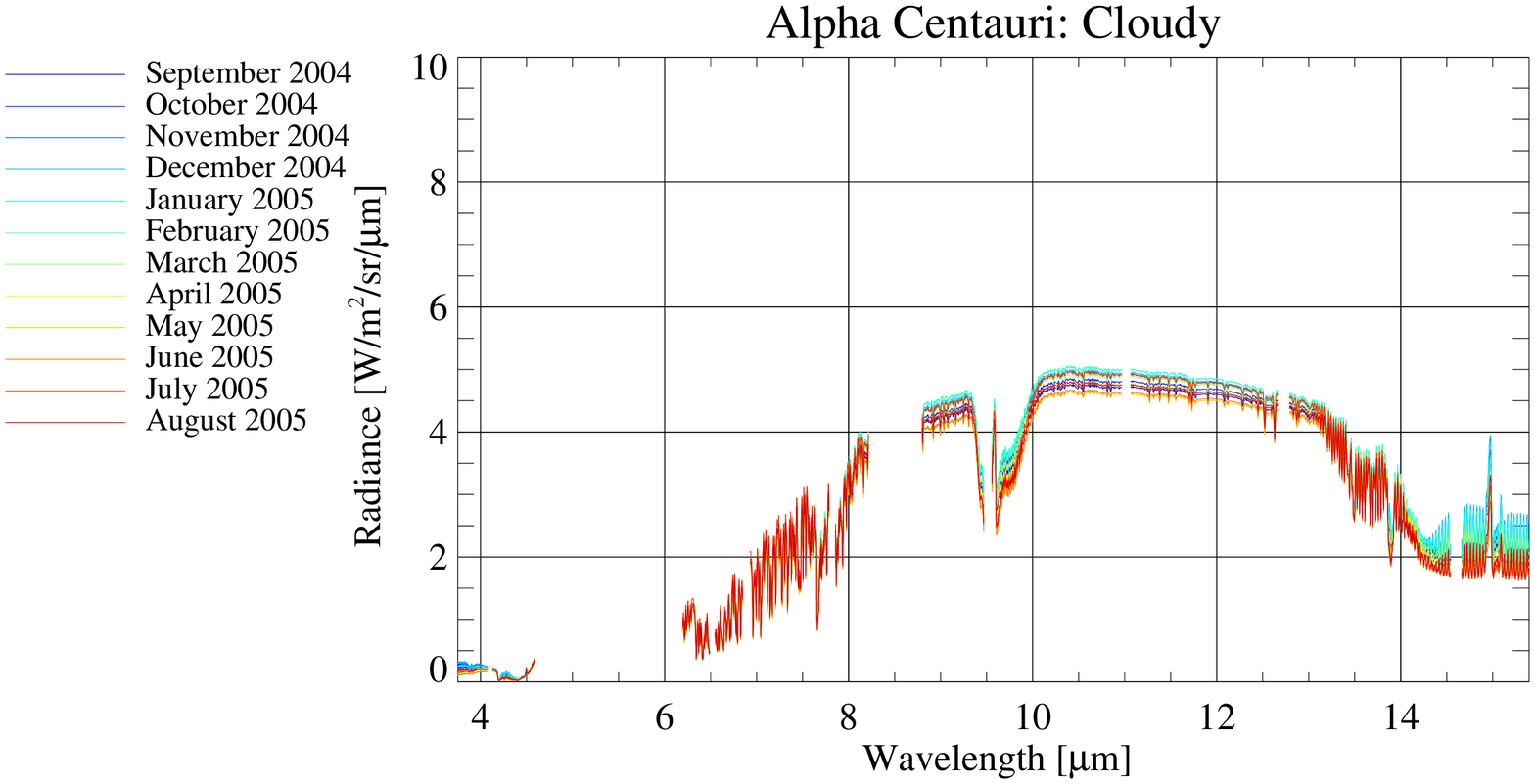}
\end{center}
\caption{The Seasonal variations are displayed for 3 pole-on views of Earth for Clear, Normal, and Cloudy cases.}
\label{fig:ir_polar_seasons}
\end{figure}

Figure ~\ref{fig:ir_polar_seasons} shows the seasonal variations in the mid-infrared for a view of Earth
from RA = 14.6 hours \& Dec = $-$60.63$^{\circ}$, the ``$\alpha$ Centauri'' view.
The major spectral features in the mid infrared are visible for all seasons
but the H$_2$O lines are weaker for the more cloudy cases.
Since this nearly pole-on view is shifted to the southern hemisphere the warmest
spectrum is in the austral Summer.  However, since there is less land in the southern
hemisphere, the amplitude of the seasonal radiometric variability is less than for views that
see the northern hemisphere.

\clearpage
\pagestyle{empty}
\setlength{\voffset}{-12mm}

\clearpage
\pagestyle{plaintop}
\setlength{\voffset}{0mm}

\section{Conclusions}
\label{conclusion}

Cloud amount and patchiness will be the most observable characteristic of extrasolar Earth-like
planets in the mid-infrared since clouds can alter the apparent temperature and the detectability
of mid-infrared spectral features that are tracers for habitability and life.
Viewing geometry can have a similar effect.  Molecular lines of H$_2$O, CH$_4$, CO$_2$, and O$_3$
are detectable in the disk averaged spectra of all geometries and cloud amounts examined.
However, he robustness of the detections degrades at lower spectral resolution particularly
for cloudier cases.

If an extrasolar planet is completely covered with cloud, O$_3$ and CO$_2$ may
not appear in absorption even if they are present.  However, the lines may appear
in emission if they are present above the cloud layer and observed with sufficient spectral
resolution (R $\sim$ 100).

The AIRS observations confirm that rotational and seasonal variations in the mid-infrared
can be as large as $\sim$ 10\%
for a planet with a tilted rotation axis and an uneven distribution of land and ocean.
However, the amplitude of this variability may be reduced for some viewing geometries
and cloud coverage.  The rotational variations in the infrared and visible
may be used to infer the period of the planet.

\acknowledgments

An anonymous referee provided several suggestions which made this paper more relevant
to missions aimed at detecting extrasolar terrestrial planets.
Conversations with Victoria Meadows helped plan the
scope of this research.  TH wishes to acknowledge Tom Pagano and Moustafa Chahine for support
in the early stages of this work and Joel Susskind for support in the later stages.
John Gieselman provided computer support without which this analysis would not have
been possible.  Some of this work was carried out at the Jet Propulsion Laboratory,
California Institute of Technology, under contract with NASA.  Some of the results in this
paper have been derived using the HEALPix \citep{gor05} package.

\end{document}